\documentclass[%
 reprint,
superscriptaddress,
nofootinbib,
amsmath,amssymb,
aps,
twocolumn,
nobalancelastpage
]{revtex4-2}
\usepackage[utf8]{inputenc}

\usepackage{graphicx}% Include figure files
\usepackage{dcolumn}% Align table columns on decimal point
\usepackage{bm}% bold math
\usepackage{xcolor}
\usepackage{adjustbox}
\usepackage{amsthm}
\usepackage{amsmath}
\usepackage{amssymb}
\usepackage{multirow}
\usepackage{hyperref}
\usepackage{physics}
\usepackage{siunitx}
\usepackage{natbib}
\usepackage{appendix}

\newcommand{\UCY}{Department of Physics, University of Cyprus, P.O. Box 20537, 1678 Nicosia, Cyprus}
\newcommand{\CYI}{Computation-based Science and Technology Research Center, The Cyprus Institute, Nicosia, Cyprus}
\newcommand{\Wuppertal}{University of Wuppertal, Wuppertal, Germany}

\begin{document}

\title{Low-lying baryon masses using twisted mass fermions ensembles at the physical pion mass}
\author{Constantia Alexandrou}
\affiliation{\UCY}
\affiliation{\CYI}
\author{Simone Bacchio}
\affiliation{\CYI}
\author{Georgios Christou}
\affiliation{\UCY}
\author{Jacob Finkenrath}
\affiliation{\CYI}
\affiliation{\Wuppertal}

\date{\today}

\begin{abstract}

We investigate the low-lying baryon spectrum using three  $N_f = 2 + 1 + 1$ ensembles simulated with physical values of the quark masses and lattice spacings of  $0.080,\; 0.069\text{ and } 0.057$ fm. The ensembles are generated using twisted mass clover-improved fermions and the Iwasaki gauge action. The spatial length is kept approximately the same at about  5.1~fm to 5.5~fm fulfilling the condition $m_\pi L> 3.6$. We investigate isospin splitting within isospin multiples and verify that for most cases the isospin splitting for these lattice spacing is consistent with zero. In the couple of cases, for which there is a non-zero value, in the continuum limit, the mass splitting goes to zero.  
The baryon masses are extrapolated to the continuum limit using the three $N_f=2+1+1$ ensembles and are compared to other recent lattice QCD results. For the strange and charm quark masses we find, respectively, $m_s(2~\rm{GeV})=99.2(2.7)$~MeV and $m_c(3~{\rm GeV)}=1.015(39)$~GeV.
The values predicted for the masses of the doubly charmed $\Xi_{cc}^\star$, $\Omega_{cc}$ and $\Omega_{cc}^\star$  baryons are 3.676(55)~GeV, 3.703(51)~GeV and 3.803(50)~GeV, respectively, and for the triply charmed $\Omega_{ccc}$ baryon is 4.785(71)~GeV. 
\end{abstract}

\maketitle

\section{Introduction}
Baryon spectroscopy is an active field of research.  Experimentally, new baryons are being studied at LHCb and CERN, by the SELEX and FOCUS  experiments at Fermilab, the BaBaR experiment at SLAC,  and the Belle-II at KEK. The discovery of new baryons like the doubly charmed $\Xi$ resonance by SELEX and LHCb triggered a revival
of the interest in charmed baryon spectroscopy.  The Beijing Spectrometer BES-III has an extensive program in spectroscopy and so does PANDA planned at GSI~\cite{PANDA:2009yku}. 

There are also many theoretical studies in baryon spectroscopy using QCD sum rules~\cite{Wang:2010hs},  as well as relativistic \cite{Martynenko:2007je, Ebert:2002ig} and non-relativistic quark models \cite{Roberts:2007ni}.  Recent progress in algorithms and access to larger computational resources have led to tremendous progress in simulating lattice QCD at physical values of the quark masses. Therefore, lattice QCD is in a good position to investigate the masses of doubly and triply charmed baryons using simulations with physical values of the quark masses providing valuable input directly from first principles. While baryon masses of low-lying hyperons and singly charmed baryons are well determined experimentally and thus serve as benchmark quantities, the doubly and triply charmed
sector still remain mostly unexplored.  A number of lattice QCD groups have studied the ground states of spin-1/2 and spin-3/2 charmed baryons using a variety of lattice schemes, with the most recent ones using dynamical simulations~\cite{Na:2007pv, Na:2008hz, Briceno:2012wt, Liu:2009jc, Basak:2012py, Durr:2012dw, PACS-CS:2013vie, Brown:2014ena, Perez-Rubio:2015zqb}. However, these calculations still performed a chiral extrapolation since ensembles with multiple lattice spacings at the physical point are limited. 

In this work, we use three ensembles generated by the Extended Twisted Mass Collaboration (ETMC) with two degenerate twisted mass clover-improved light quarks with mass tuned to reproduce the physical pion mass, plus a strange and charm quark with masses fixed to approximately their physical ones. We will refer to these $N_f=2+1+1$ ensembles as physical point ensembles. These physical point ensembles have lattice spacings $a=0.08$~fm, 0.069~fm, and 0.057~fm and approximately equal volume. This enables us, for the first time, to take the continuum limit directly at the physical point, thus eliminating a systematic uncertainty arising from the chiral extrapolation that in the baryon sector can introduce an uncontrolled systematic error. Including a clover-term helps in the stabilization of the simulations, while it still preserves automatic $\mathcal{O}(a)$-improvement of the twisted mass action and reduces the $\mathcal{O}(a^2)$ lattice artifacts related to the breaking of the isospin symmetry.  This study extends our previous computations on the low-lying baryon spectrum using $N_f = 2$ with one physical point ensemble~\cite{Alexandrou:2017xwd}.

\begin{table*}[t!]
        \centering
        \begin{tabular}{c|c|c|c|c|c|c|c|c|c}
                Ensemble & $V$ & $\beta$ & $\mu_\ell$ & $\mu_\sigma$ & $\mu_\delta$ & $\kappa$ & $c_{SW}$& $L\cdot m_\pi$ & $m_\pi[\si{MeV}]$\\
                \hline\hline
                cB211.072.64 & $128\times 64^3$ & 1.778 & 0.00072 & 0.1246826 & 0.1315052 & 0.1394265 & 1.69&3.62 &  140.1 (0.2) \\
                cC211.060.80 & $160\times 80^3$ & 1.836 & 0.00060 & 0.106586 & 0.107146 & 0.13875285 & 1.6452&3.78 & 136.7 (0.2) \\
                cD211.054.96 & $192\times 96^3$ & 1.900 & 0.00054 & 0.087911 & 0.086224 & 0.137972174 & 1.6112&3.90 & 140.8 (0.2)\\
        \end{tabular}
        \caption{
   Simulation parameters used by ETMC for generating each ensemble~\cite{Finkenrath:2022eon}. The first column gives the ensemble name, the second the lattice volume, the third the $\beta$-value, and the fourth, fifth, and sixth the quark mass parameters. In the seventh and eighth columns, we give the critical value of $\kappa$  for which $m_l\sim 0$ and the value of $c_{SW}$. In the last but one column we give $L m_\pi$ and in the last column the pion mass.}
        \label{tab:Ensemble_Information}
\end{table*}

\begin{table}[t!]
    \begin{tabular}{c|c|c|c|c}
        Ensemble& $a_G$ & $n_G$ & $a_\text{APE}$ & $n_\text{APE}$\\
        \hline\hline
        cB211.072.64 & 1.0 & 95 & 0.5& 50\\
        cC211.060.80 & 1.0 & 140 & 0.5& 60\\
        cD211.054.96 & 1.0 & 200 & 0.5& 60\\
    \end{tabular}
    \caption{Smearing parameters used for all ensembles.}
    \label{tab:gparams}
\end{table}

We follow our previous studies and use Osterwalder-Seiler (OS) valence strange and charm quarks and choose the physical masses of the $\Omega^-$ and the $\Lambda_c^+$   baryons to tune the OS valence strange and charm quark masses, respectively. We also use the nucleon mass to fix the lattice spacing in order to convert our lattice values to physical units. 
Isospin breaking in the twisted mass formulation is a lattice artifact of order ${a^2}$. It has been shown that adding the clover term reduces isospin splitting in the $\Delta$-multiplet~\cite{ETM:2015ned} as compared to the $N_f = 2 + 1 + 1$ twisted mass simulations at a similar lattice spacing. Here we study the effects of isospin breaking effects to higher accuracy for the $\Delta$ multiplet and in the strange and charm sectors. We also extract the renormalized strange and charm quark masses and find agreement with our previous determination~\cite{ExtendedTwistedMass:2021gbo}. We compare our final results on the masses of the forty baryons studied in this work with those of other recent lattice calculations, using a variety of discretization schemes, as well as with experiments. We find agreement with experimental results. Our predictions for the doubly charmed $\Xi_{cc}^\star$, $\Omega_{cc}$ and $\Omega_{cc}^\star$  baryons are $M_{\Xi_{cc}^\star} = 3.676(55)\text{GeV}$, $M_{\Omega_{cc}} = 3.703(51)\text{GeV}$ and $M_{\Omega_{cc}^\star} = 3.803(50)\text{GeV}$, respectively, and for the triply charmed $\Omega_{ccc}$ baryon is $M_{\Omega_{ccc}} = 4.785(71)\text{GeV}$.

The paper is organized as follows:  the lattice action employed in this work, as well as the details of the calculations, including the interpolating fields and the effective mass analysis procedure, are given in Section II. The determination of the lattice spacing and the tuning of the strange and charm quark masses are given in Sections III and IV respectively, and we present our lattice results in Section V.  In Section~\label{sec:isospin}, we study the isospin splitting of the spin-1/2 and spin-3/2 baryons.  In Section VII we present our results for the baryon spectrum at the continuum limit and in Section VIII a comparison between our results and other lattice QCD results is presented. Lastly, we present our conclusions in Section IX.

\section{Lattice setup}
\subsection{The lattice action}
We analyze three gauge ensembles produced by ETMC~\cite{Alexandrou:2018egz, Finkenrath:2022eon, Finkenrath:2023sjg} within the twisted mass fermion formulation~\cite{Frezzotti:1999vv}, whose parameters are given in Table~\ref{tab:Ensemble_Information}. The $N_f$ = 2 + 1 + 1 ensembles are generated using the Iwasaki gauge action,  the $N_f$= 2 mass-degenerate twisted mass fermion action for the light doublet with a clover term ~\cite{Sheikholeslami:1985ij} and the $N_f$ = 1 + 1 non-degenerate twisted mass fermion action for the strange and charm quarks. Including a clover term, reduces lattice artifacts and decreases the mass gap between the charged and neutral pion.  We use 1-loop tadpole boosted perturbation theory to fix the value of the clover parameter $c_{\rm SW}$. The light quark mass is tuned to reproduce the isosymmetric pion mass $m_\pi=0.135$~MeV, the strange quark mass is tuned to reproduce the physical value of the ratio between $D_s$ meson mass and decay constant, and the charm quark is fixed by setting the ratio of charm to strange quark mass $\mu_c/\mu_s=11.8$~\cite{Finkenrath:2022eon}.

\begin{table}[t!]
\centering
        \begin{tabular}{c|c}
                Ensemble & $\mu^\prime_\ell$\\
                \hline\hline
                cB211.072.64 & 0.0006675\\
                cD211.054.96 & 0.0004964\\
        \end{tabular}
        \caption{Light quark twisted masses $\mu_\ell$ used for correcting the effective mass~\cite{ExtendedTwistedMass:2022jpw}.}
        \label{tab:Pion_Correction_Parameters}
\end{table}

For completeness, we provide the action used in the simulations. The Iwasaki gauge action is given by  
\begin{equation}
    \begin{split}
        S_g & = \frac{\beta}{3}\sum_x\left(b_0\sum_{\mu<\nu}\left[1 - \Re \trace(U_{\mu\nu}^{1\times 1}(x))\right]\right.\\
        & \; \; \; \; \; \; \; \; \;\; \; \; \;\;\;\;\; \left.+ b_1\sum_{\mu<\nu}\left[1 - \Re\trace(U_{\mu\nu}^{1\times 2}(x))\right]\right),
    \end{split}
\end{equation}
where  $\beta = 6/g_0^2$  is the inverse bare coupling constant and we set the parameters $b_0 = 1 - 8b_1$ and $b_1 = -0.331$ such that the Iwasaki improved gauge action \cite{Iwasaki:1983iya} is reproduced.

\begin{figure*}[t!]
        \centering
        \includegraphics[width=0.95\linewidth]{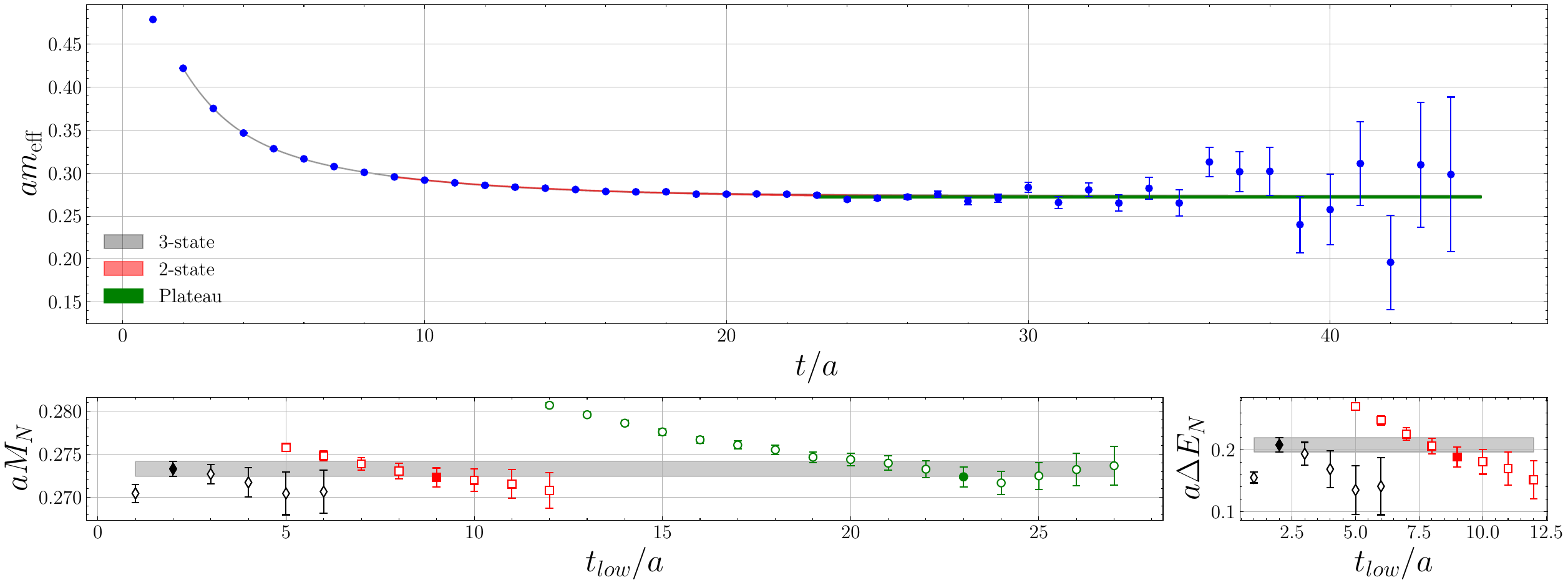}
        \caption{The effective mass for the nucleon using the cD211.054.96 ensemble. On the top panel, we present one- (green band), two- (red band) and three-state (grey band) fits for which the extracted nucleon mass is consistent. On the bottom left panel, we show the value extracted from the plateau fit (green points) as a function of the value of $t_{\rm low}$ used in the fit, while, on the bottom right panel, we show the energy difference $\Delta E$ between the first excited state and the ground state extracted from the two- (red) and three-state (black) fits.  On the bottom panels, the filled symbols give the selected values for the ground state (bottom left) and $\Delta E$ (bottom right), and the error band is centered around the result of the most probable model. All values are given in lattice units.}
        \label{fig:D96_Nucleon}
\end{figure*}

\begin{table*}[t!]
\centering
               \begin{tabular}{c|c|c|c|c|c|c|c}
                        \multirow{2}{*}{Ensemble} & \multirow{2}{*}{$t_\text{max}/a$} & \multicolumn{2}{c|}{one-state} & \multicolumn{2}{c|}{two-state} & \multicolumn{2}{c}{three-state}\\
                         &  & $t_\text{low}/a$ & $aM_B$ & $t_\text{low}/a$ & $aM_B$ & $t_\text{low}/a$ & $aM_B$\\
                        \hline\hline
                        cB211.072.64 & 34 & 18 & 0.3819(18) & 8 & 0.3796(24) & 3 & 0.3798(34)\\
                        cC211.060.80 & 38 & 18 & 0.32756(78) & 8 & 0.3258(12) & 3 & 0.3257(17)\\
                        cD211.054.96 & 44 & 23 & 0.2724(11) & 9 & 0.2723(11) & 2 & 0.27330(84)\\
                \end{tabular}
        \caption{Nucleon effective mass results from the most probable one-, two- and three-state fits.
        For each fit, we give in lattice units the maximum and minimum times $t_{\rm maax}$ and $t_{\rm low}$ used in the fit, as well as the baryon mass extracted from each fit.}
        \label{tab:Nucleon_Fit_Results}
    \end{table*}

For the fermionic sector, we distinguish between light- $(u,d)$ and heavy-flavored $(c,s)$ quarks.
We construct the light and heavy quark doublets $\chi_\ell = (u, d)^T$ and $\chi_h = (s, c)^T$ in the twisted basis, respectively. The light quark action reads
\begin{equation}
    \begin{split}
        S_{tm}^\ell & = \sum_x \bar{\chi}_\ell(x)\left[D_W[U] + \frac{i}{4}c_{SW}\sigma_{\mu\nu}\mathcal{F}^{\mu\nu} + m_\ell\right.\\
        & \;\;\;\;\;\;\;\;\;\;\;\;\;\;\;\;\;\;\;\;\;\;\;\; + \left.\frac{}{}i\mu_\ell\gamma_5\tau^3\right]\chi_\ell(x),
    \end{split}
\end{equation}
and the heavy quark part of the action reads
\begin{equation}
    \begin{split}
        S_{tm}^h & = \sum_x \bar{\chi}_h(x)\left[D_W[U] + \frac{i}{4}c_{SW}\sigma_{\mu\nu}\mathcal{F}^{\mu\nu} + m_h\right.\\
        & \;\;\;\;\;\;\;\;\;\;\;\;\;\;\;\;\;\;\;\;\;\;\;\; -\mu_\delta \tau^1  + \left.\frac{}{}i\mu_\sigma\gamma_5\tau^3\right]\chi_h(x).
    \end{split}
\end{equation}
The third Pauli matrix $\tau^3$ acts in flavor space, $D_W$ is the massless Wilson-Dirac operator and $\mu_l$ is the light quark twisted mass. Automatic  $\mathcal{O}(a^2)$ improvement is achieved by tuning $m_l$ to zero using the partially conserved axial current (PCAC) relation. 
 In the heavy quark action, the additional term $\mu_\delta \tau^1$ is acting on flavor space, thus, the mass degeneracy, in contrast to the light quark sector, is lifted.
The massless Wilson-Dirac operator is given by
\begin{equation}
    D_W[U] = \frac{1}{2}\gamma_\mu\left(\nabla_\mu + \nabla_\mu^\star\right) - \frac{ar}{2}\nabla_\mu\nabla_\mu^\star,
\end{equation}
where the forward derivatives read
\begin{equation}
    \nabla_\mu\psi(x) = \frac{1}{a}\left[U_\mu^\dagger(x)\psi(x + a\hat{\mu} - \psi(x)\right]
\end{equation}
and 
\begin{equation}
    \nabla_\mu^\star\psi(x) = -\frac{1}{a}\left[U_\mu(x - a\hat{\mu})\psi(x - a\hat{\mu} - \psi(x)\right].
\end{equation}
At or close to maximal twist, we transform from twisted to physical basis using
$\psi(x) = \frac{1}{\sqrt{2}}\left[1+ i\tau^3\gamma_5\right]\chi(x)$ and 
$\bar{\psi}(x) = \bar{\chi}(x)\frac{1}{\sqrt{2}}\left[1 + i\tau^3\gamma_5\right]$.
Throughout this paper, unless otherwise stated, we use quark fields in the physical basis, in particular when we define baryon interpolators.
For the valence strange and charm quarks, we utilize Osterwalder Seiler (OS) fermions~\cite{Osterwalder:1977pc}. We discuss the tuning of the OS strange and charm quark masses below.

\subsection{Two-point functions and effective masses}
In this work, we consider positive parity baryon correlators of two-point functions of the form
\begin{equation}
    C_B^\pm(t,\vec{0}) = \sum_{\vec{x}_f}\expval{\frac{1}{4}\trace(1\pm\gamma_0)\mathcal{J}_B(\vec{x}_f,t_f)\bar{\mathcal{J}}_B(\vec{x}_i,t_i)}\,,%e^{-i\vec{p}\cdot\vec{x}_f},
\end{equation}
where $\bar{\mathcal{J}}_B$ is the interpolating field of the baryon acting at the source $(\vec{x}_i,t_i)$ and ${\mathcal{J}}_B$ at the sink $(\vec{x}_f,t_f)$.
We consider three-quark baryon interpolating fields of the form $\varepsilon_{abc}\left(q_{1, a}^T\Gamma q_{2, b}\right)q_{3, c}$.
The $\Gamma$ structures we use are $C\gamma_5$ and $C\gamma_i, i = 1,\ldots,3$ for spin-$1/2$ and $3/2$, respectively.

We exploit the symmetries of the action and the anti-periodic boundary conditions in the temporal direction to write
\begin{equation}
    C^+_B(t, \vec{0}) = -C^-_B(T-t, \vec{0}).
\end{equation}
Thus, we average the correlators in the forward and backward direction to improve the signal by taking 
\begin{equation}
    C_B(t, \vec{0}) = C^+_B(t, \vec{0}) - C^-_B(T - t, \vec{0}).
\end{equation}
For the spin-3/2 baryons,  we project to spin $3/2$  by performing the following transformations at zero momentum~\cite{PhysRevC.39.2339}
\begin{equation}
        C_{\frac{3}{2}}(t) = \frac{1}{3}\trace(C(t)) + \frac{1}{6}\sum_{i\neq j}^{3}\gamma_i\gamma_j C_{ij}(t),
\end{equation}
where $t=t_f-t_i$ and from now on we drop the momentum dependence in the argument of the two-point function correlator since we only consider $\vec{p}=\vec{0}$.

Since, in this work, we are interested in the low-lying spectrum and gauge noise grows with the time $t$, we need to use techniques that suppress excited state contamination. We apply Gaussian smearing to the quark fields at the source and sink \cite{Gusken:1989qx, Alexandrou:1992ti} using
\begin{equation}
        q_\text{smear}(n) = \sum_{m\in\Lambda}F(n, m, U)q(m),
\end{equation}
where the gauge invariant smearing function reads
\begin{equation}
        F(n, m, U) = (1 + \alpha_GH)^{n_G}(n, m, U)
\end{equation}
and $H$ is the hopping function
\begin{equation}
        H(n, m, U) = \sum_{j = 1}^3\left(U_j(n)\delta_{n + a\hat{j}, m} + U^\dagger_j(n-a\hat{j})\delta_{n - a\hat{j}, m}\right).
\end{equation}
The parameters $a_G$ and $n_G$ for the three ensembles are given in Table~\ref{tab:gparams}. We also apply APE smearing to the links that enter the hopping function $H(n,m,U)$ with APE smearing parameter $a_\text{APE}$ and number of iterations $n_\text{APE}$ for all ensembles given in Table~\ref{tab:gparams}.

To extract the low-lying masses, we construct the effective mass using  correlators at zero momentum
\begin{equation}
    am_{\text{eff}}^B(t) = \log(\frac{C(t)}{C(t + 1)}) \xrightarrow[t\to\infty]{}aM_B.
\end{equation}
We adopt a multi-state fit approach to identify the ground state keeping up to the second excited state.
 Namely, we fit the effective mass to a constant, and to the following expressions taking into account the first and second excited states, respectively
\begin{equation}
        \label{eq:two_state_fit}
        am_{\text{eff}}^B(t) \approx aM_B + \log(\frac{1 + \alpha_2e^{-\Delta E_1 t}}{1 + \alpha_2e^{-\Delta E_1 (t + 1)}})
\end{equation}
and
\begin{equation}
        \label{eq:three_state_fit}
        am_{\text{eff}}^B(t) \approx aM_B + \log(\frac{1 + \alpha_2e^{-\Delta E_1 t} + \alpha_3 e^{-\Delta E_2 t}}{1 + \alpha_2e^{-\Delta E_1 (t + 1)} + \alpha_3 e^{-\Delta E_2 (t + 1)}}).
\end{equation}
We fix the maximum time used in fit to given value $t_{\rm max}$ and subsequently, we iteratively change the time from where we start the fit $t_{\rm low}$ for all three types of fits. We refer to the different fit ranges as models.

\section{Lattice spacing determination}

\begin{table}[t!]
\centering
        \begin{tabular}{c|c|c}
                & $a$~[fm] & $a$~[fm]\\
                \hline
                Ensembles & AIC Selection & Model Averaging\\
                \hline\hline
                cB211.072.64 & 0.07982(51) & 0.07998(52)\\
                cC211.060.80 & 0.06852(24) & 0.06853(27)\\
                cD211.054.96 & 0.05747(18) & 0.05719(43)\\
        \end{tabular}
        \caption{The values of the lattice spacing for the three ensembles using the maximum probability fit (second column) and averaging over all models (third column).}
        \label{tab:Ensemble_Spacings}
\end{table}

\begin{table}[t!]
        \begin{tabular}{c|c|c|c}
                Ensemble & $\mu_s$ & $\mu_c$ & $Z_p$\\
                \hline\hline
                \multirow{2}{*}{cB211.072.64} & 0.017 & 0.20 & \multirow{2}{*}{0.4788(34)(42)}\\
                & 0.019 & 0.22& \\
                \hline
                \multirow{2}{*}{cC211.060.80} & 0.015 & 0.18 & \multirow{2}{*}{0.4871(26)(42)}\\
                & 0.017 & 0.24& \\
                \hline
                \multirow{2}{*}{cD211.054.96} & 0.0125 & 0.146 & \multirow{2}{*}{0.4967(7)(36)}\\
                & 0.0152 & 0.18 & \\
        \end{tabular}
        \caption{The values of the two bare quark masses used for the strange (second column) and charm (third column) quark flavors for the interpolation to the tuned value. In the fourth column, we give the renormalization constants \cite{ExtendedTwistedMass:2021gbo} $Z_P$  in the $\overline{\text{MS}}$ scheme at $2\,\si{GeV}$ for each ensemble, with the value in the first and second parenthesis being the statistical and systematic error, respectively.}
        \label{tab:Bare_Quark_Masses}
\end{table}

For the tuning of the lattice spacing, we opt to use the physical nucleon mass as input.
From the value of $m_\pi$ given in Table~\ref{tab:Ensemble_Information}, it can be seen that two ensembles have slightly higher pion mass as compared to the one target by our simulations, which is $135$~MeV.  We correct for this small deviation by  
generating at lower statistics two-point functions for lighter twisted masses with values shown in Table~\ref{tab:Pion_Correction_Parameters} and applying a correction to the effective mass as follows
\begin{equation}
        \label{eq:Effective_Mass_Correction}
        \begin{split}
                a\delta m_\text{eff}^B(t) & = \log(\frac{C(t)}{C(t + 1)}\frac{C^\prime(t + 1)}{C^\prime(t)})\xrightarrow[t\to\infty]{}a\Delta M_B,\\
        \end{split}
\end{equation}
where $C^\prime(t)$ denotes the correlator evaluated at the lower value of the twisted mass parameter.

The corrected nucleon effective mass reads
\begin{equation}
        \label{eq:Corrected_Effective_Mass}
        \begin{split}
                am_\text{eff}^{\prime B}(t) & = am_\text{eff}^B(t) + a\delta m_\text{eff}^B(t)\\
                & = \log(\frac{C(t)}{C(t + 1)}) - \log(\frac{C(t)}{C(t + 1)}\frac{C^\prime(t + 1)}{C^\prime(t)}).
        \end{split}
\end{equation}
This correction changes the nucleon mass on a percent level and ensures that all baryon masses computed with these three ensembles correspond to the same physical pion mass.

\begin{figure*}[t]
        \centering
        \includegraphics[width=0.95\linewidth]{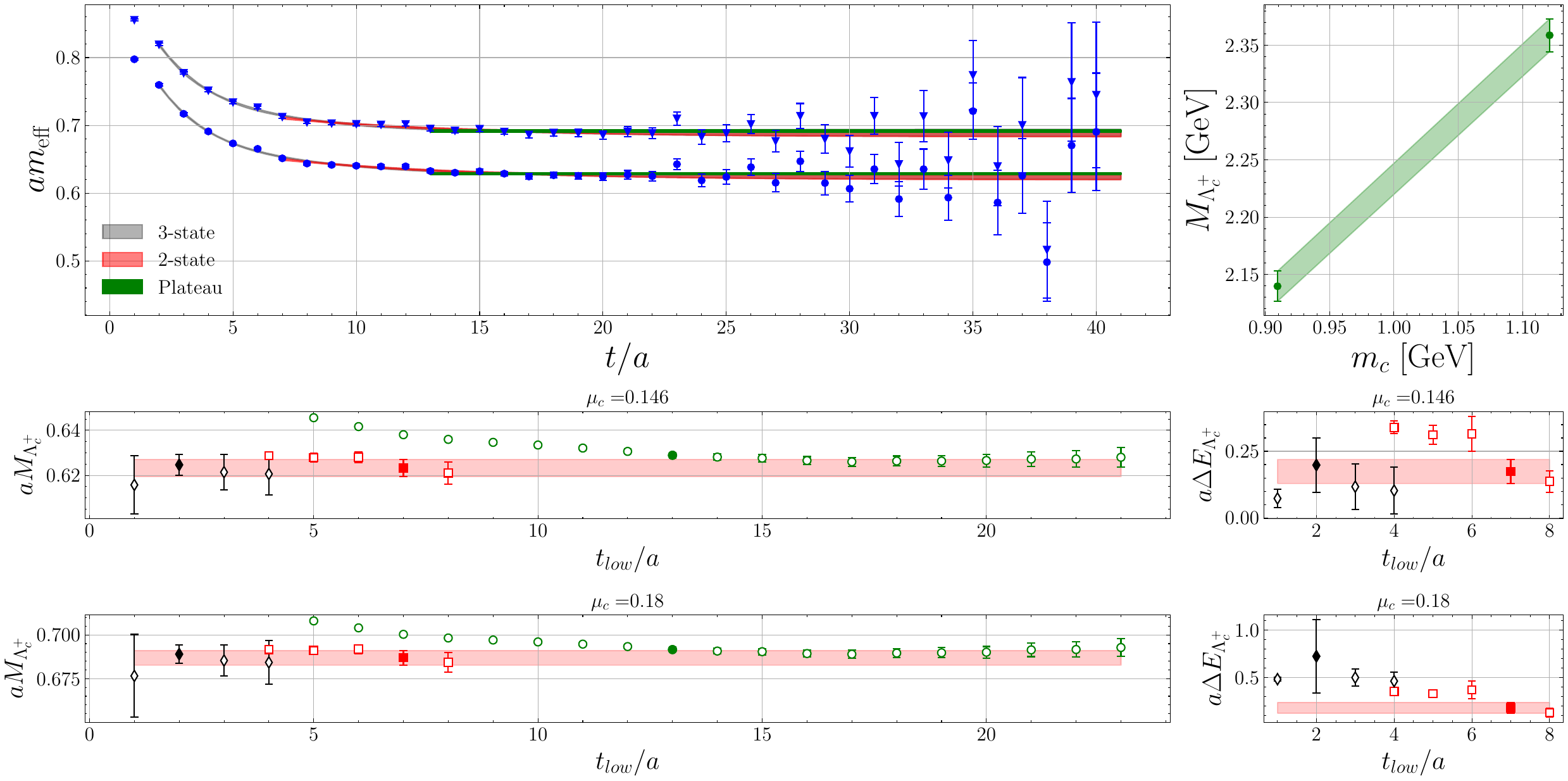}
        \includegraphics[width=0.95\linewidth]{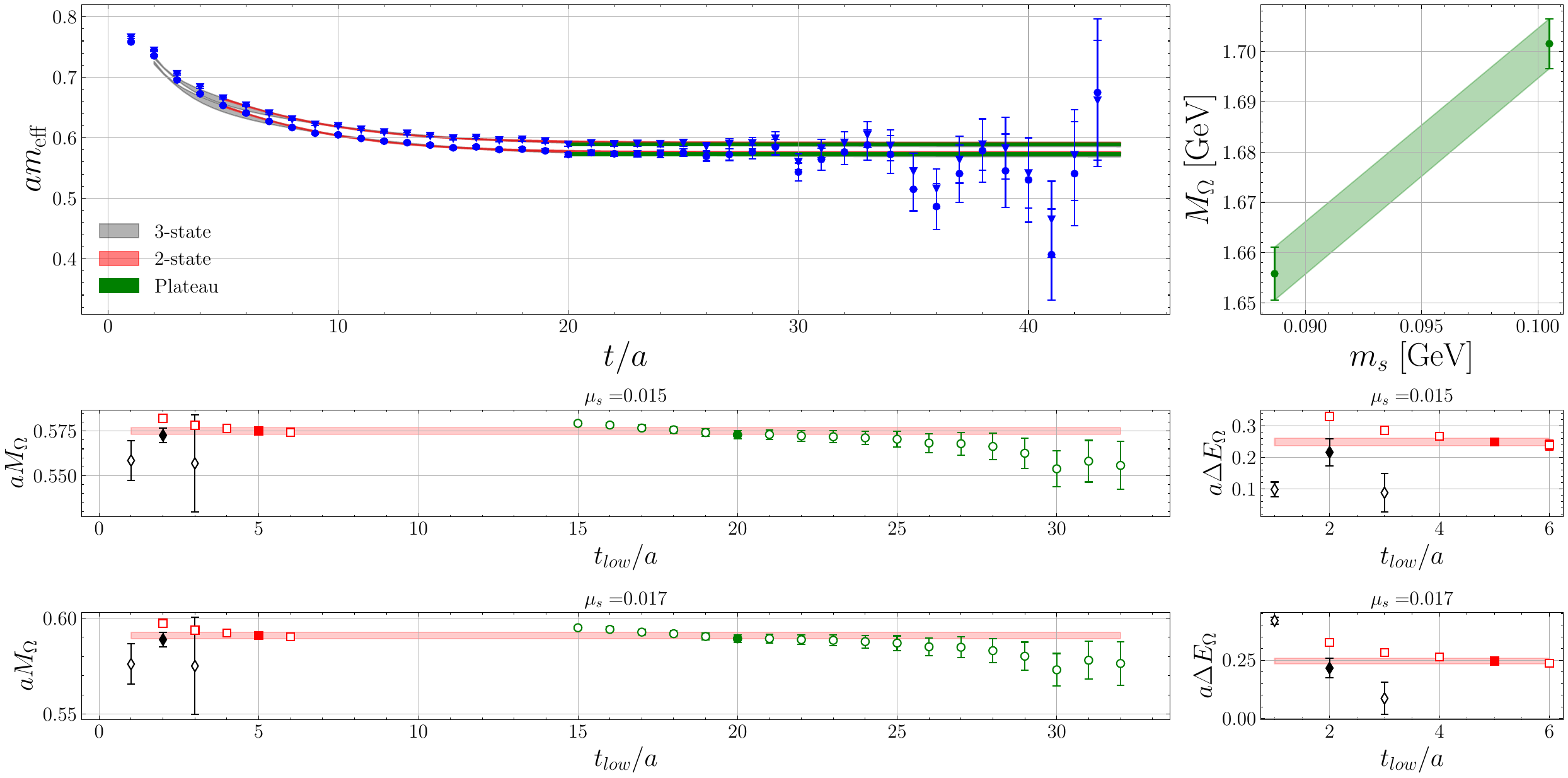}
        \caption{We show the effective mass analysis for the $\Lambda_c^+$ (top) and $\Omega^-$ (bottom) baryon for the cD211.054.96 and cC211.060.80 ensembles respectively.
        On the top left we depict the effective mass data for the two values of $\mu_c$ as well as the one-, two-, and three-state fits. On the top left we show the linear interpolation.
        On the middle and bottom panels, we show the results for the mass for the two $\mu_c$ values and the difference between the ground and first excited state extracted from each fit.
        The green and red bands represent the most probable model result. The rest of the notation is the same as that of Fig.~\ref{fig:D96_Nucleon}.}
        \label{fig:LambdacPlD96}
\end{figure*}

\begin{table*}[t!]
        \begin{tabular}{c|c|c|c|c|c|c|c|c|c|c}
                \multicolumn{11}{c}{Tuning of the strange quark mass }\\
                \hline\hline
                \multirow{2}{*}{Ensemble} & \multirow{2}{*}{$a\mu_s$} & \multirow{2}{*}{$t_{\text{max}}/a$} & \multicolumn{2}{c|}{one-state}& \multicolumn{2}{c|}{two-state} & \multicolumn{2}{c|}{three-state} & \multirow{2}{*}{$A_{\Omega^-}\si{GeV}$} & \multirow{2}{*}{$B_{\Omega^-}$} \\
                 &  &  & $t_\text{low}/a$ & $aM_B$ & $t_\text{low}/a$ & $aM_B$ & $t_\text{low}/a$ & $aM_B$ &  & \\
                \hline\hline
                \multirow{2}{*}{cB211.072.64} & 0.017 & 35 & 14 & 0.6768(12) & 5 & 0.6741(16) & 3 & 0.6729(22) & \multirow{2}{*}{1.6948(38)} & \multirow{2}{*}{3.957(48)} \\
                 & 0.019 & 35 & 14 & 0.6932(11) & 5 & 0.6906(15) & 3 & 0.6899(19) & \\
                \hline
                \multirow{2}{*}{cC211.060.80} & 0.015 & 43 & 20 & 0.5729(23) & 5 & 0.5750(18) & 2 & 0.5726(41) & \multirow{2}{*}{1.6803(51)} & \multirow{2}{*}{3.866(50)} \\
                 & 0.017 & 43 & 20 & 0.5893(21) & 5 & 0.5909(17) & 2 & 0.5888(37) & & \\
                \hline
                \multirow{2}{*}{cD211.054.96} & 0.0125 & 52 & 23 & 0.4784(14) & 7 & 0.4774(14) & 2 & 0.4774(17) & \multirow{2}{*}{1.6772(46)} & \multirow{2}{*}{3.894(48)} \\
                 & 0.0152 & 52 & 23 & 0.4997(12) & 7 & 0.4985(13) & 2 & 0.4984(16) & & \\
                \hline
                \multicolumn{11}{c}{Charm quark mass evaluation}\\
                \hline\hline
                \multirow{2}{*}{Ensemble} & \multirow{2}{*}{$a\mu_c$} & \multirow{2}{*}{$t_{\text{max}}/a$} & \multicolumn{2}{c|}{one-state}& \multicolumn{2}{c|}{two-state} & \multicolumn{2}{c|}{three-state} & \multirow{2}{*}{$A_{\Lambda_c^+}\si{GeV}$} & \multirow{2}{*}{$B_{\Lambda_c^+}$} \\
                 &  &  & $t_\text{low}/a$ & $aM_B$ & $t_\text{low}/a$ & $aM_B$ & $t_\text{low}/a$ & $aM_B$ &  & \\
                \hline\hline
                \multirow{2}{*}{cB211.072.64} & 0.20 & 30 & 12 & 0.8663(21) & 5 & 0.8631(35) & 1 & 0.8598(98) & \multirow{2}{*}{2.389(11)} & \multirow{2}{*}{0.9463(99)} \\
                 & 0.22 & 30 & 12 & 0.9020(22) & 5 & 0.8987(38) & 1 & 0.895(12) & & \\
                \hline
                \multirow{2}{*}{cC211.060.80} & 0.18 & 31 & 12 & 0.7584(20) & 3 & 0.7579(22) & 3 & 0.757(14) & \multirow{2}{*}{2.4104(74)} & \multirow{2}{*}{0.9456(59)} \\
                 & 0.24 & 31 & 12 & 0.8633(24) & 3 & 0.8629(27) & 3 & 0.862(15) & & \\
                \hline
                \multirow{2}{*}{cD211.054.96} & 0.146 & 40 & 13 & 0.6289(13) & 7 & 0.6232(38) & 2 & 0.6247(46) & \multirow{2}{*}{2.445(15)} & \multirow{2}{*}{1.034(12)} \\
                 & 0.18 & 40 & 13 & 0.6916(14) & 7 & 0.6870(42) & 2 & 0.6890(54) & & \\
        \end{tabular}
\caption{We give the $\Omega^-$ (top) and $\Lambda_c^+$(bottom) baryon mass results from the most probable one-, two- and three-state fits. For each fit we give $t_{\rm min}$ and $t_{\rm max}$ and the interpolation parameters from Eq.~\eqref{eq:2conf_interpolation} as determined from  the most probable model.}
\label{tab:OmegaMn_LambdacPl_Fits}
\end{table*}

    \begin{table*}[t!]
    \centering
        \begin{tabular}{c|c|c|c|c}
                \multirow{2}{*}{Ensemble} & \multicolumn{2}{c|}{Strange quark} & \multicolumn{2}{c}{Charm quark}\\
                & AIC Selection & Model Averaging & AIC Selection & Model Averaging\\
                \hline\hline
                cB211.072.64 & 0.0894(10) & 0.0894(11) & 1.092(10) & 1.091(19) \\
                cC211.060.80 & 0.0930(13)  & 0.0931(14) & 1.0689(72)& 1.0708(93)\\
                cD211.054.96 & 0.0938(12)  & 0.0939(13) & 1.047(13)& 1.050(17)\\
        \end{tabular}
        \caption{Tuned renormalized quark masses for each ensemble  in $\si{GeV}$ in the $\overline{\text{MS}}$ scheme. The strange quark masses are given at $2\,\si{GeV}$ and the charm at $3\,\si{GeV}$.}
        \label{tab:Ensemble_Tuned_Masses}
\end{table*}

For the determination of the lattice spacing we implement the following analysis method:
We use the Akaike Information Criterion (AIC)~\cite{Jay:2020jkz, Neil:2022joj} and calculate the model probability for each fit function and for each fit range
\begin{eqnarray}
        \label{eq:Akaike}
        P(M_i | \{y\}) &=& \frac{1}{Z}\exp(-\frac{1}{2}\chi_i^2 + N_{\text{d.o.f}, i}), \nonumber\\  Z &= &\sum_i P(M_i|\{y\}),
\end{eqnarray}
where $\chi_i^2$ is the chi-squared and $N_{\text{d.o.f}, i}$ is the number of degrees of freedom used for the specific model.
Subsequently, we select the model with the highest probability and fix the nucleon mass to its physical value. 
Another approach is to average over all models and fix the nucleon mass using the model average (MA) value
\begin{eqnarray}
        \expval{\alpha} &=& \sum_i \alpha_i P(M_i | \{y\}),  \nonumber\\ \sigma_\alpha^2 &=& \sum_i (\alpha_i^2 + \sigma_{\alpha, i}^2)P(M_i | \{y\}) - \expval{\alpha}^2.
\end{eqnarray}
We show the nucleon effective mass and results of various fits in Fig.~\ref{fig:D96_Nucleon}. In Table~\ref{tab:Nucleon_Fit_Results}, we give the values of the nucleon mass extracted from the plateau, two- and three-state fits for the case of the fits for the largest probability.

The values of the lattice spacings that we extract using the nucleon mass as input are given in Table~\ref{tab:Ensemble_Spacings}. 
Since both approaches employed are consistent with each other we select the MA result as our final values of the lattice spacing since they take into account systematic errors in the variation of the fit ranges.

\section{Determination of the strange and charm quark masses}

For tuning the valence strange and charm quark masses we use the $\Omega^-$ and $\Lambda_c^+$ baryon masses respectively.
In order to tune the quark masses we generate two sets of correlators for two different values of the non-renormalized strange and charm quark masses. The values used are listed in Tab.~\ref{tab:Bare_Quark_Masses}. Since the data are correlated two sets suffice per ensemble to perform the linear interpolation.
In Fig. \ref{fig:LambdacPlD96}, we illustrate the analysis of the effective mass for $\Lambda_c^+$ and $\Omega^-$. The procedure is analogous to the extraction of the nucleon mass. The most probable one-, two- and three-state fit results for $\Omega-$ and $\Lambda_c^+$ baryons are given in Tab~\ref{tab:OmegaMn_LambdacPl_Fits}.

We implement three different methods to perform the quark mass tuning.
In method-I, we first renormalize the quark masses and interpolate to the physical baryon masses for each ensemble using  
\begin{eqnarray}
        \label{eq:2conf_interpolation}
        M_\Omega &= A_{\Omega} + B_{\Omega}(m_s - \tilde{m}_s), \nonumber\\
         M_{\Lambda_c} &= A_{\Lambda_c} + B_{\Lambda_c}(m_c - \tilde{m}_c),
\end{eqnarray}
  where we take $\tilde{m}_s=0.095\,\si{GeV}$ and $\tilde{m}_c=1.2\,\si{GeV}$.
This is done for all combinations of fits ranges, i.e. for all models.
Subsequently, using Eq.~\eqref{eq:Akaike} we select the model with the highest probability as our final result.
Alternatively, we can average over all models with the corresponding probabilities and calculate a model average of the tuned quark masses. The strange and charm quark masses calculated using method-I are given in Tab.~\ref{tab:Ensemble_Tuned_Masses}.

\begin{figure}[t!]
    \centering
    \includegraphics[width = 0.79\linewidth]{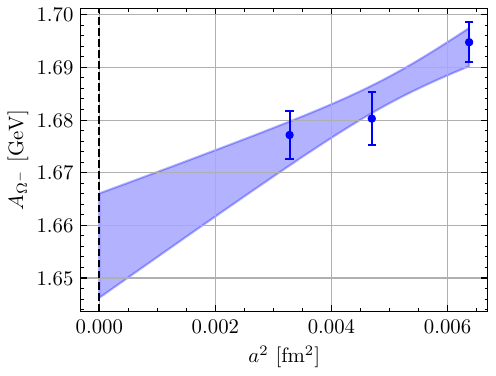}
    \includegraphics[width = 0.79\linewidth]{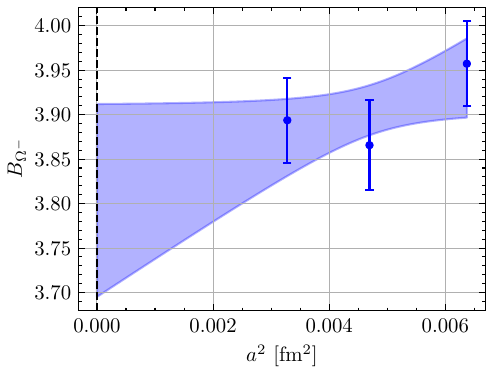}
    \vspace*{-0.2cm}
    \caption{Continuum extrapolation of the interpolation parameters as described in method-II for $\Omega^-$.}
    \label{fig:Method_III_Omega}
    \vspace{0.2cm}
    \centering
    \includegraphics[width = 0.79\linewidth]{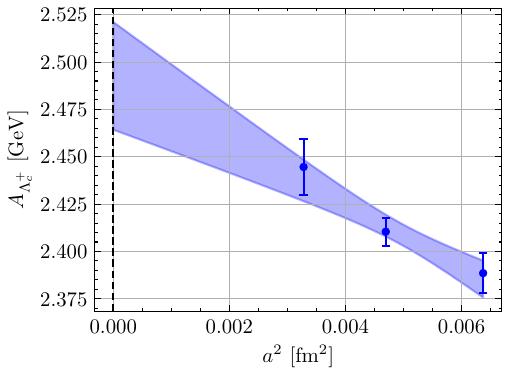}
    \includegraphics[width = 0.79\linewidth]{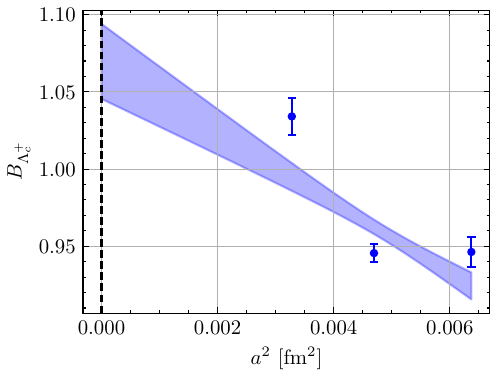}
    \vspace*{-0.2cm}
    \caption{Continuum extrapolation of the interpolation parameters as described in method-II for $\Lambda_c^+$.}
    \label{fig:Method_III_Lambda}
\end{figure}
\begin{figure}[t!]
    \centering
    \includegraphics[width = 0.79\linewidth]{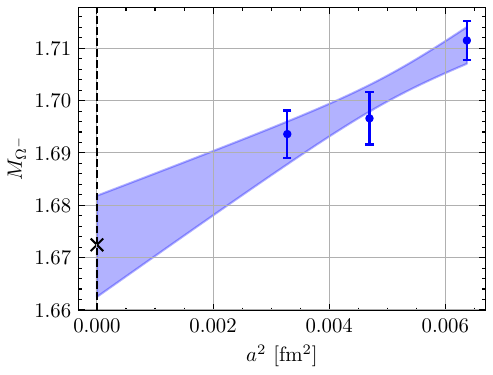}
    \includegraphics[width = 0.79\linewidth]{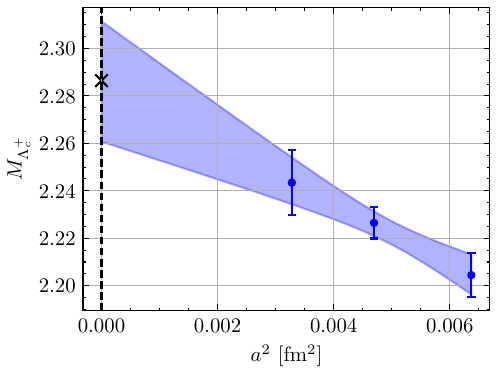}
    \caption{The continuum limit extrapolation for the mass of the $\Omega^-$ (top) and the $\Lambda_c^+$ (bottom) used for method-III after tuning the mass of the strange and charm quarks. The cross represents the physical value of their mass.}
    \label{fig:Method-IV}
\end{figure}
\begin{table}
\centering
        \begin{tabular}{c|c|c}
                 & $m_s(2~{\rm GeV})$ & $m_c(3~(\rm{GeV})$\\
                \hline\hline
                Method-I (AIC)                  & 0.0991(26) & 1.006(27)\\
                Method-I (MA)                   & 0.0992(27) & 1.015(39)\\
                Method-II                       & 0.0994(26) & 1.012(24)\\
                Method-III                      & 0.0994(26) & 1.011(24)\\
        \end{tabular}
        \caption{We give the values of the renormalized strange and charm quark masses at the continuum limit in $\si{GeV}$, using all three methods. Both values are given in the $\overline{\rm MS}$ scheme at a scale of 2~GeV and 3~GeV, respectively for the strange and charm quarks.}
        \label{tab:qmasses_all_methods}
\end{table}

In method-II, we perform the interpolation for the results of the most probable model using Eq.~\eqref{eq:2conf_interpolation} and then extrapolate the interpolation parameters to the continuum limit using
\begin{eqnarray}
        A_{\Omega^-, \Lambda_c^+}(a) = c_{\Omega^-, \Lambda_c^+} + d_{\Omega^-, \Lambda_c^+}a^2, \nonumber\\
        B_{\Omega^-, \Lambda_c^+}(a) = e_{\Omega^-, \Lambda_c^+} + f_{\Omega^-, \Lambda_c^+}a^2.
\end{eqnarray}
 The continuum extrapolation of the parameters $A_{\Omega^-, \Lambda_c^+}$ and $B_{\Omega^-, \Lambda_c^+}$ are shown in Fig.~\ref{fig:Method_III_Omega} and \ref{fig:Method_III_Lambda}.
Subsequently, we calculate the strange quark mass quark mass using
\begin{equation}
        m_{s, c} = \tilde{m} + \frac{M_{\Omega^-, \Lambda_c^+}^\text{phys} - A_{s, c}(0)}{B_{s, c}(0)},
\end{equation}
where $M^{\text{phys}}_\Omega$ is the physical value of $\Omega^-$. We repeat the procedure for the tuning of the charm quark using the mass of $\Lambda_c$. 

\begin{figure*}[t!]
    \includegraphics[width =0.95\linewidth]{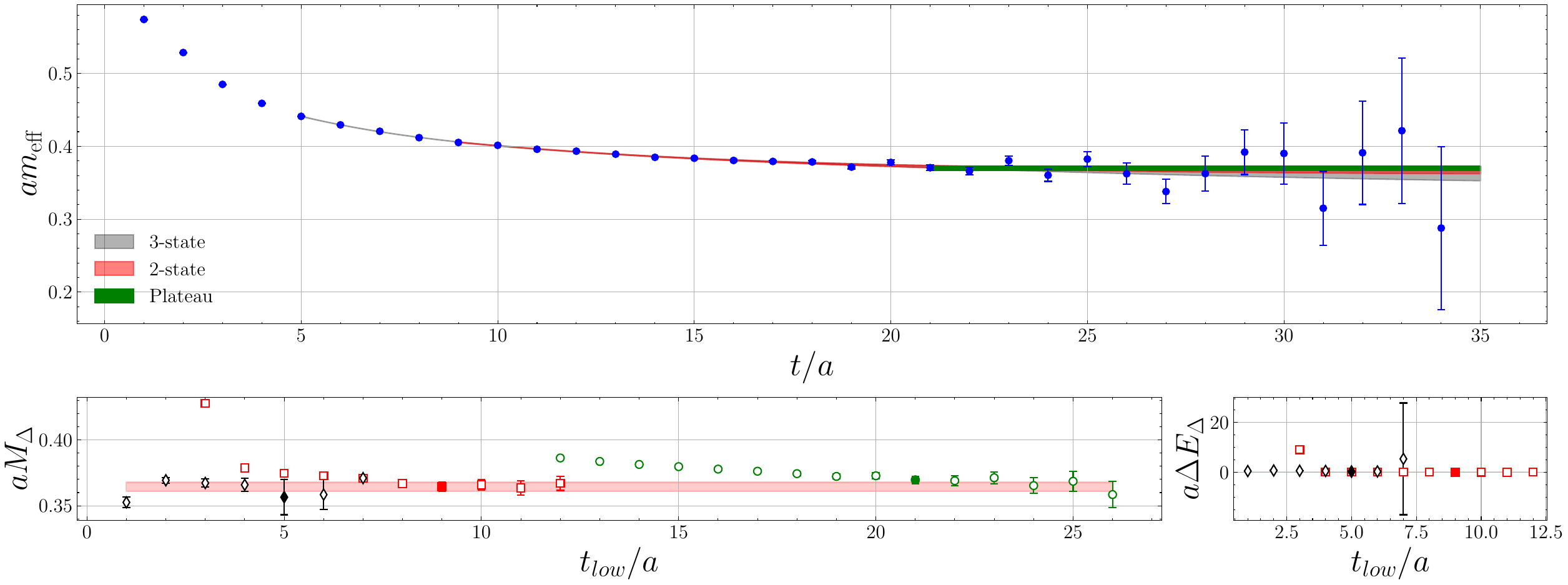}
    \caption{$\Delta$ baryon effective mass analysis. We use the same notation as Fig. \ref{fig:D96_Nucleon}.}
    \label{fig:delta}
\end{figure*}

\begin{figure*}[t!]
    \centering
    \includegraphics[width = \textwidth]{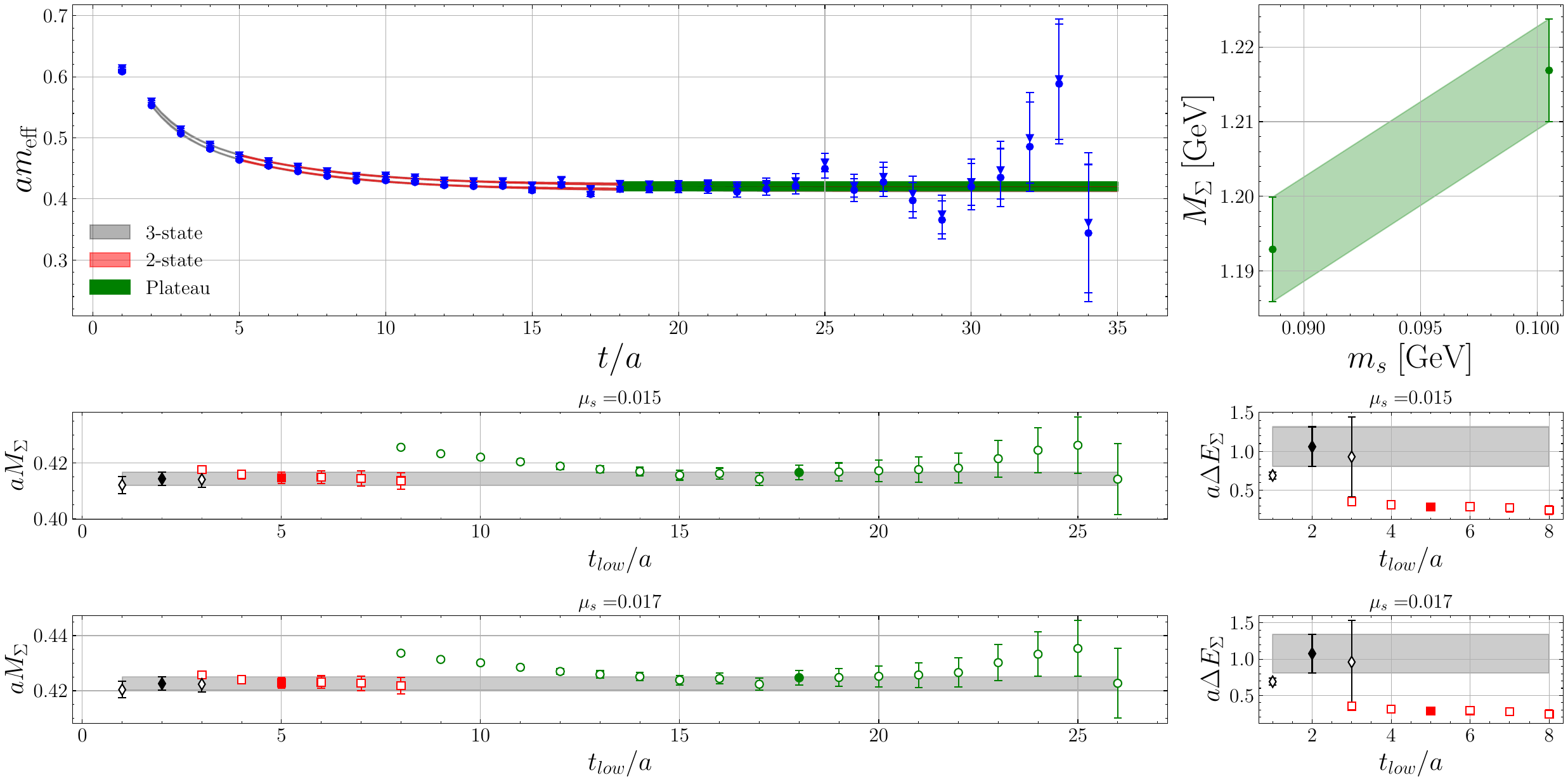}
    \includegraphics[width = \textwidth]{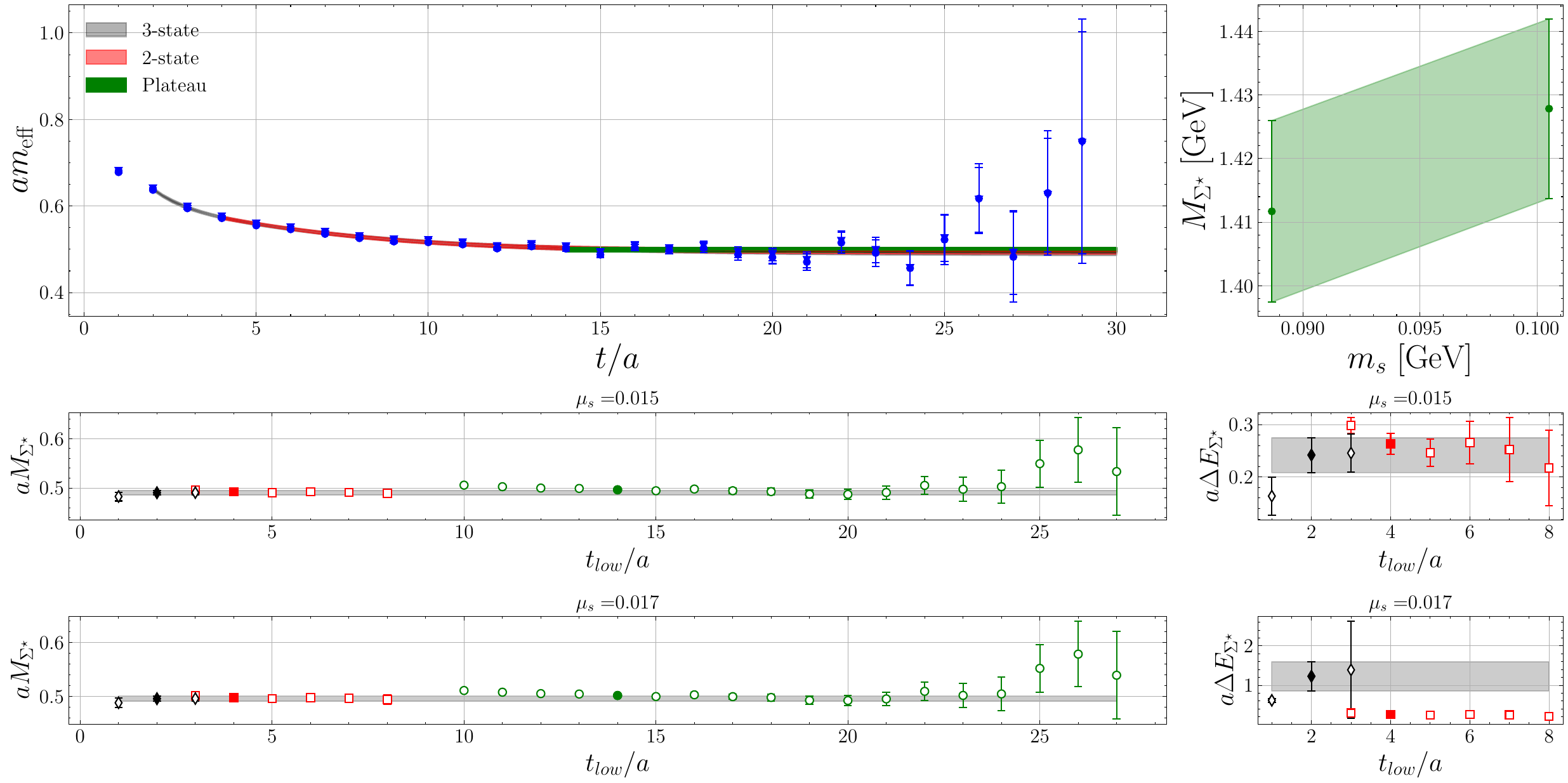}
    \caption{Results on the effective mass of non-charmed baryon containing a strange quark. From top to bottom for the $\Sigma$, and $\Sigma^\star$  baryons computed using the cC211.060.80 ensemble. The notation is the same as that in Fig. \ref{fig:LambdacPlD96}.}
    \label{fig:sigma}
\end{figure*}

\begin{figure*}[t!]
    \centering
    \includegraphics[width = \textwidth]{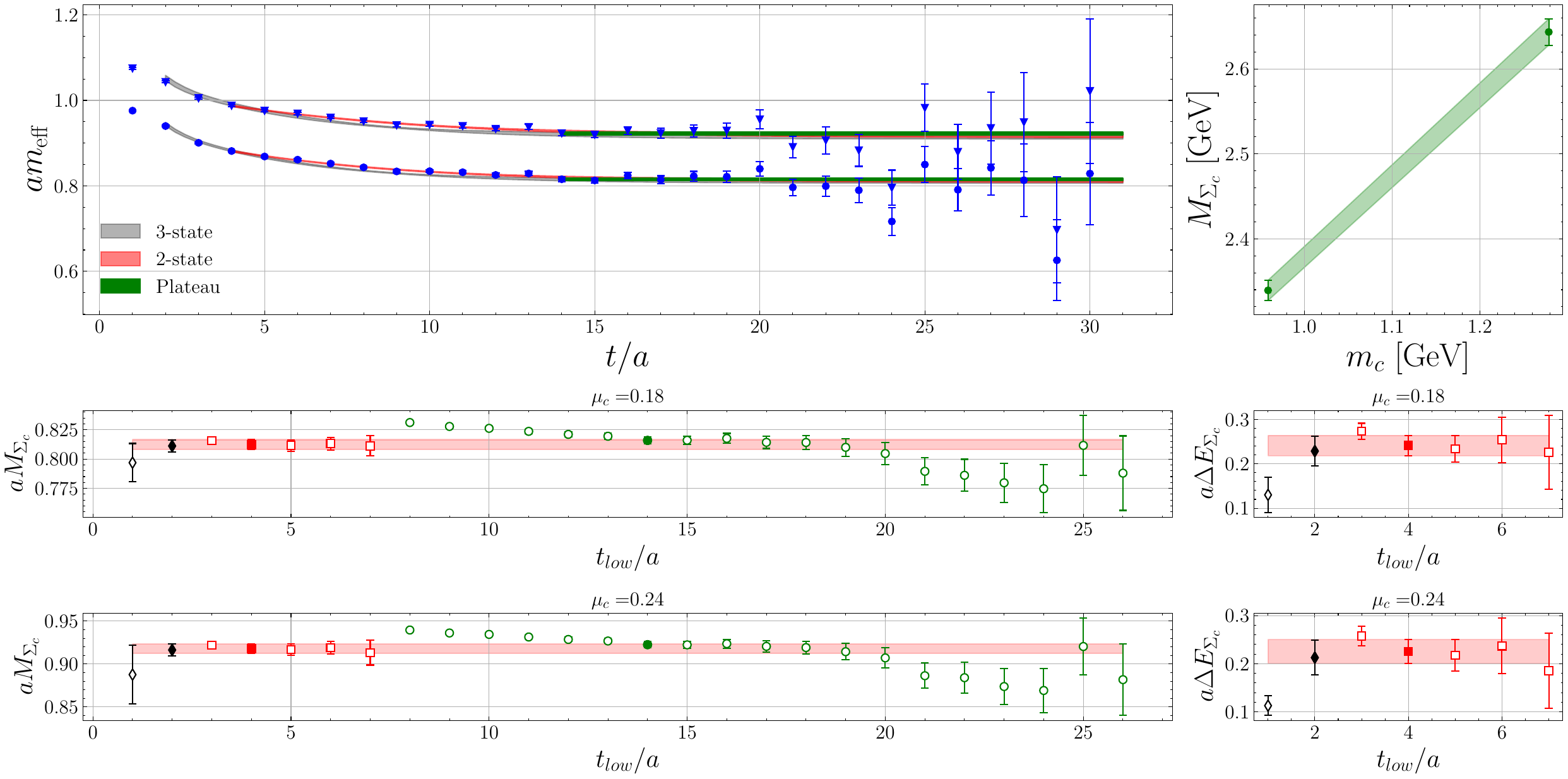}
    \includegraphics[width = \textwidth]{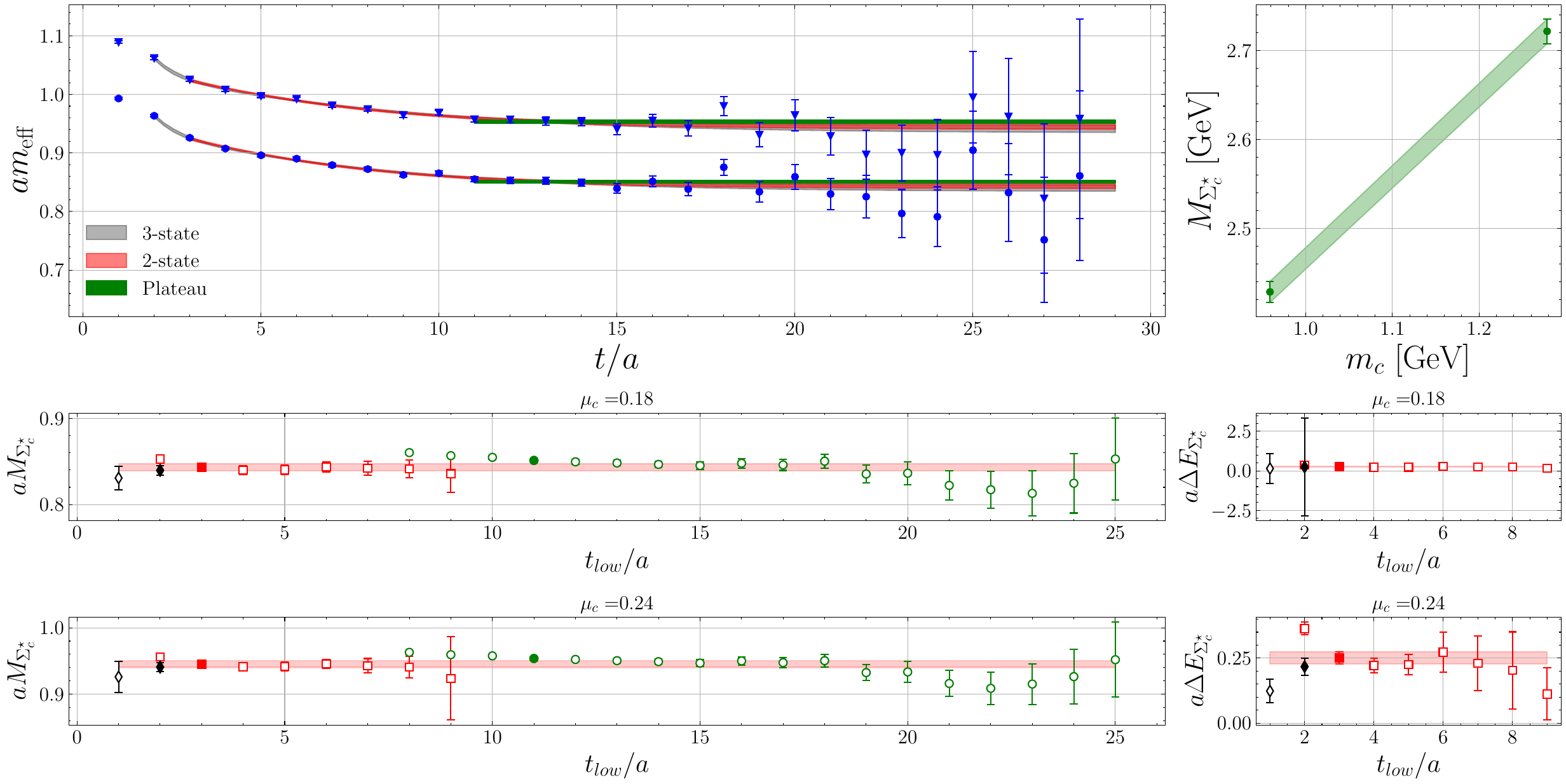}
    \caption{Results on the effective mass of charmed baryon containing a strange quark.  From top to bottom, we show results for the $\Sigma_c$, and $\Sigma^\star_c$  baryons computed using the cC211.060.80 ensemble. The notation is the same as that in Fig. \ref{fig:LambdacPlD96}.}
    \label{fig:sigmac}
\end{figure*}

In method-III, we adopt an iterative strategy.
Namely,  we consider a given value of the renormalized strange and charm quark mass for all ensembles and use Eq.~\eqref{eq:2conf_interpolation} to find the $\Omega^-$ and $\Lambda_c^+$ masses that correspond to this given quark masses.
We perform the continuum extrapolation of the $\Omega^-$ and $\Lambda_c^+$ masses linearly in $a^2$ and iteratively change the values of the renormalized strange and charm quark masses until the continuum limit of the $\Omega^-$ and $\Lambda_c^+$ masses match their physical values.
We present the results of method-III in Fig. \ref{fig:Method-IV}. 

We present the results on the renormalized $m_s$ and $m_c$ in the continuum limit extracted from all methods in Tab~\ref{tab:qmasses_all_methods}. As can be seen, all procedures yield the same values. We take as our final values the ones from method-I with the model average, namely
\begin{equation}
\begin{aligned}
m_s(2~{\rm GeV})&=99.2(2.7)~{\rm MeV} \quad{\rm and}\\ m_c(3~{\rm GeV})&=1.015(39)
~{\rm GeV},
\end{aligned}
\label{final_quark}
\end{equation}
in the $\overline{\text{MS}}$ scheme at the continuum limit.

\section{Extraction of baryon masses}
Having determined the lattice spacing and tuned the valence strange and charm quark masses,  we discuss in this section the analysis to extract the low-lying baryon spectrum.
For the $\Delta$, where we have only $u$ and $d$ valence quarks, only method-I is relevant since the sea and valence quarks are the same and there is no need to perform any interpolation. However,  as in the case of the nucleon mass, we do correct the $\Delta$ effective mass to take into account the slightly larger pion mass for the cB211.072.64 and cC211.054.96 ensembles. This correction is noticeable enough once we take the continuum limit and yield a result consistent with experiments. The procedure followed is the same as that used for correcting the nucleon effective mass. In Fig. \ref{fig:delta}, we show the effective mass for $\Delta$ and the corresponding one-, two- and three-states fits.
For baryons that contain strange and charm quarks, we employ all three methods described in the previous section. We find the baryon masses that correspond to the tuned masses of the strange and charm quarks using Eq.~\eqref{eq:2conf_interpolation} for the ones that contain only strange or only charm quarks. For baryons that contain both quark flavors, we use  
\begin{equation}
        \label{eq:4conf_interpolation}
        M_B = A + B(m_s - \tilde{m}_s) + C(m_c - \tilde{m}_c),
\end{equation}
with values of $\tilde{m}_s $ and $\tilde{m}_c$ the same as those used in Eq.~\eqref{eq:2conf_interpolation} employing all three methods.
Our analysis of the effective masses for selective cases is presented in Figs.~\ref{fig:delta}, \ref{fig:sigma}, and \ref{fig:sigmac}.

\section{Isospin splitting}\label{sec:splitting}

    \begin{figure*}
        \centering
        \hfill
        \includegraphics[width = 0.423\textwidth]{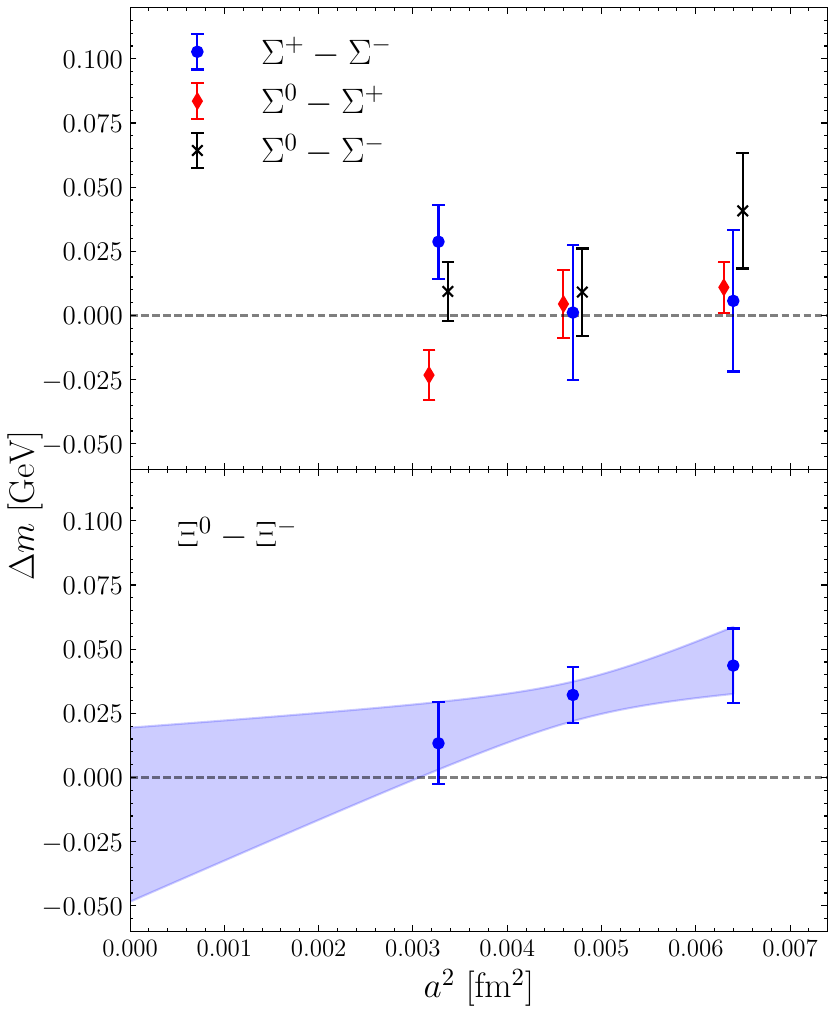}
        \hfill
        \includegraphics[width = 0.419\textwidth]{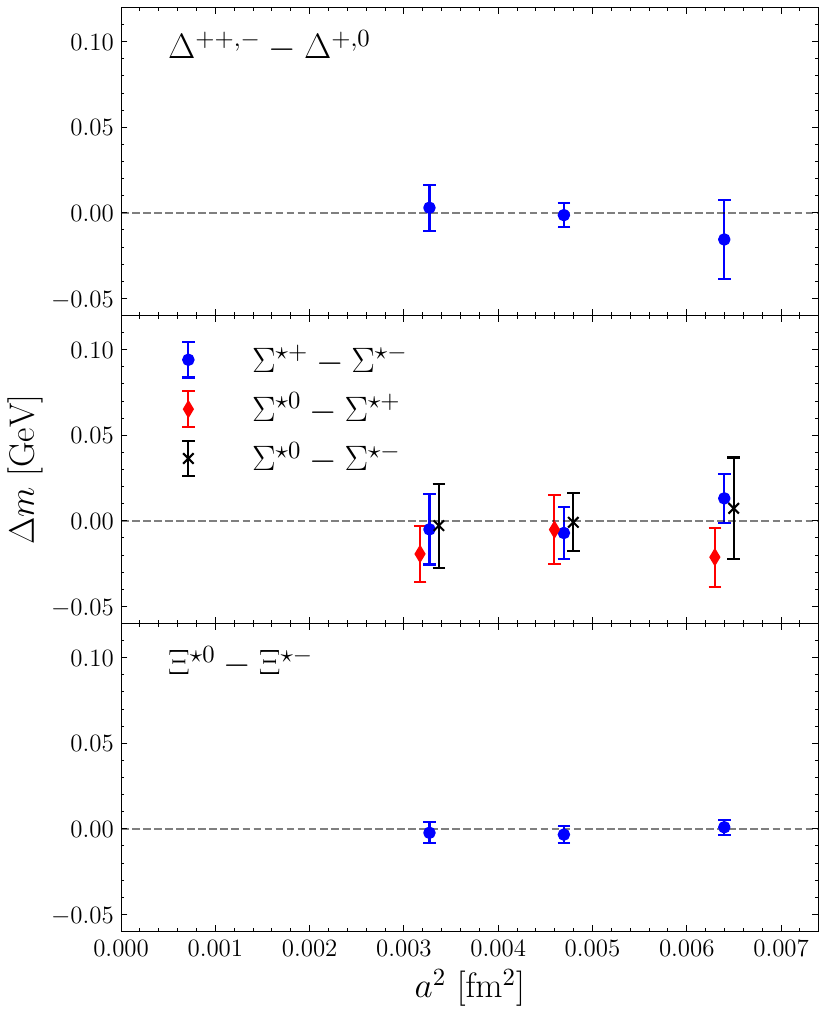}
        \hfill
        \caption{Isospin symmetry splitting for strange spin-1/2 baryons (left) and spin-3/2 baryons (right).}
        \label{fig:isospinstrange}
    \end{figure*}
    \begin{figure*}
        \centering
        \hfill
        \includegraphics[width = 0.423\textwidth]{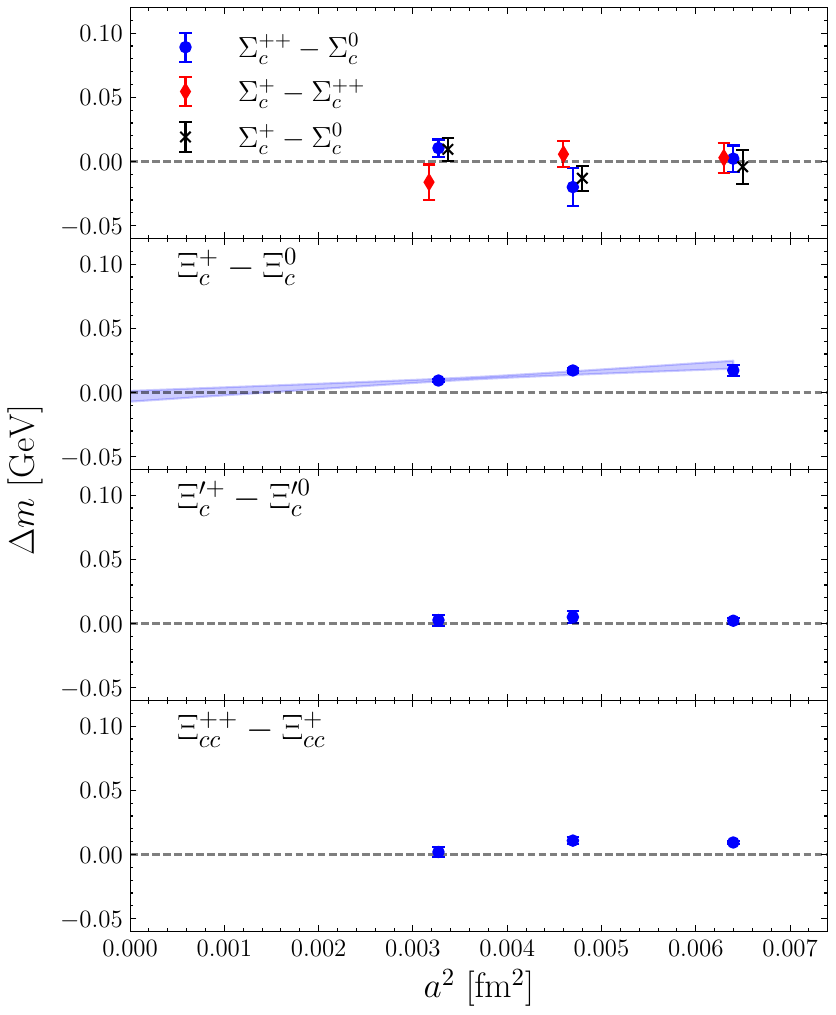}
        \hfill
        \includegraphics[width = 0.419\textwidth]{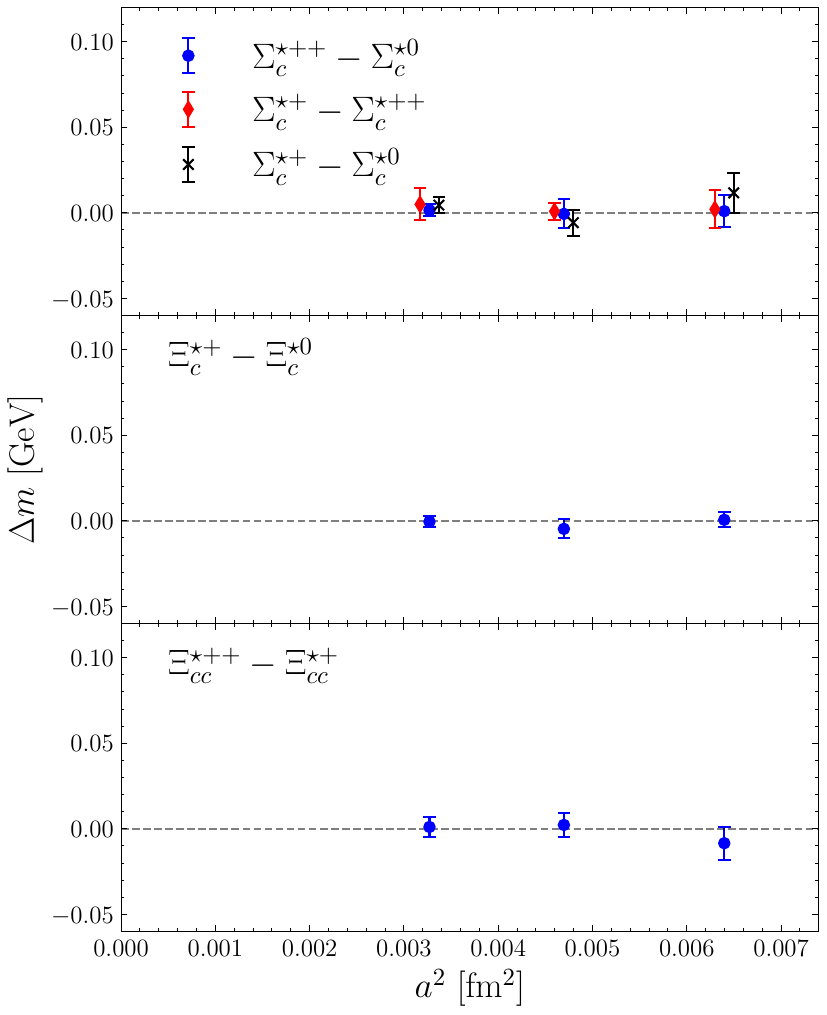}
        \hfill
        \caption{Isospin symmetry splitting for charm spin-1/2 baryons (left) and spin-3/2 baryons (right).}
        \label{fig:isospincharm}
    \end{figure*}
    
    \begin{figure}[h!]
    \centering
    \includegraphics[width = 0.77\linewidth]{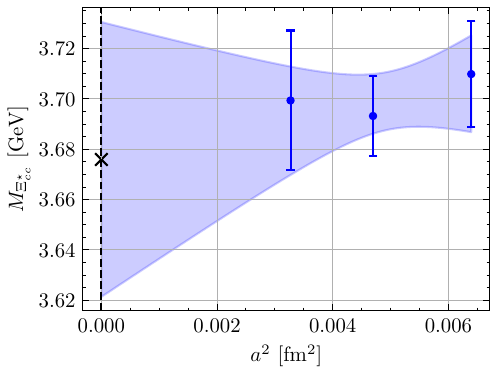}
    \includegraphics[width = 0.77\linewidth]{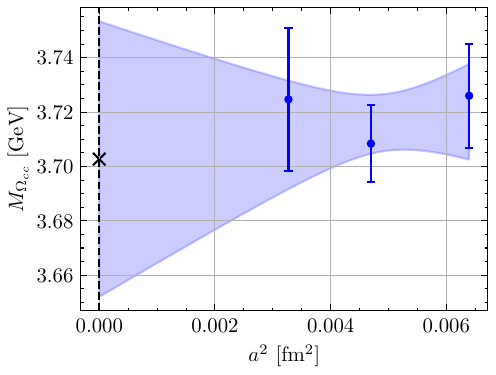}
    \includegraphics[width = 0.77\linewidth]{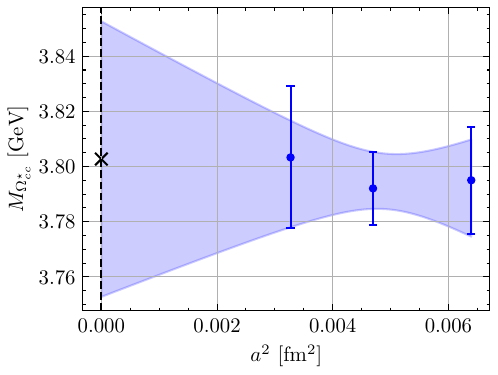}
    \vspace*{-0.2cm}
    \caption{Continuum extrapolation of the doubly charmed baryons $\Xi_{cc}^\star$, $\Omega_{cc}$ and $\Omega_{cc}^\star$ using the results of method-I with model average.}
    \label{fig:doubly_charmed}
    \centering
    \includegraphics[width = 0.77\linewidth]{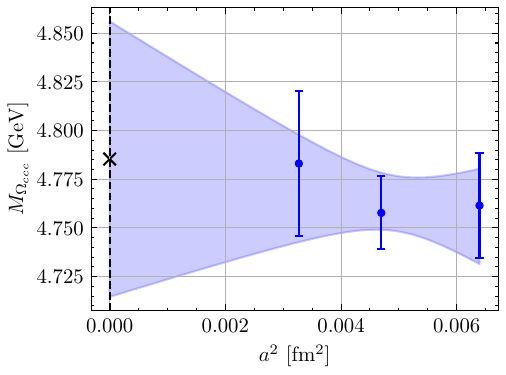}
    \vspace*{-0.2cm}
    \caption{Continuum extrapolation of the triply charmed baryon $\Omega_{ccc}$ using the results of method-I with model average.}
    \label{fig:triply_charmed}
\end{figure}

The twisted mass fermion formulation breaks isospin symmetry at finite lattice spacing. This is an ${\cal O}(a^2)$ effect and it should vanish in the continuum limit. We examine the splitting for the strange and charmed baryons as a function of the lattice spacing and in the continuum limit. 
In Figs.~\ref{fig:isospinstrange} and \ref{fig:isospincharm}, we show the results on the mass spitting for the strange and charmed baryons, respectively,  as a function of $a^2$.
As we can see, for all spin-3/2 non-charmed baryons the mass splittings are consistent with zero at all three lattice spacings. For the Delta-baryon, we average the masses of the $\Delta^{++}$ and $\Delta^-$ as well the masses of the $\Delta^-$ and $\Delta^0$, since these are degenerate by symmetry. For the spin-1/2 baryons, the only mass splitting that is not consistent with zero is $\Xi^0-\Xi^-$, which becomes zero only at the finest lattice spacing. If we extrapolate linearly to the continuum limit we see that indeed the mass splitting also in this case becomes consistent with zero.   The mass splitting of the $\Sigma^+ -\Sigma^-$ and $\Sigma^0 -\Sigma^+$ is consistent with zero for both cB211.072.64 and cC211.060.80 ensembles. The small deviation from zero for the cD211.054.96 ensemble of less than a standard deviation can, thus,  only be a statistical fluctuation since this ensemble has even smaller lattice spacing as compared to the other two. Similarly, the mass splitting between the $\Sigma^0$ and the charged $\Sigma^{-}$ is consistent with zero for all ensembles within one standard deviation.
For the charmed baryons, we also observe that all mass splittings are consistent with zero for all lattice spacings or it vanishes for the finest lattice spacing i.e. the cD211.054.96 ensemble. For the mass splitting of the $\Xi^+_c-\Xi^0_c$, that case is similar to the one of $\Xi^0-\Xi^-$ i.e.   $M_{\Xi^+_c-\Xi^0_c}$ is consistent with zero after taking the continuum limit. We include the linear extrapolation to the continuum limit in Fig.~\ref{fig:isospincharm} to highlight the fact that also this splitting becomes zero at the continuum limit.

Having shown that in the continuum limit isospin splitting vanishes we present in what follows the results on the baryons averaging among the isospin mutliplets. For the  $\Xi$ and $\Xi_c$ multiples, we average only in the continuum limit.

\section{Continuum limit}
We obtain our final values on the baryon masses by extrapolating linearly in $a^2$ to the continuum limit. We use all three methods except the case of the $\Delta$ where only method-I is relevant.
In Fig.~\ref{fig:doubly_charmed} and \ref{fig:triply_charmed} we present the continuum extrapolation for the doubly and triply charmed baryons which we predict their masses.

\begin{figure}[t!]
    \includegraphics[width = 0.9\linewidth]{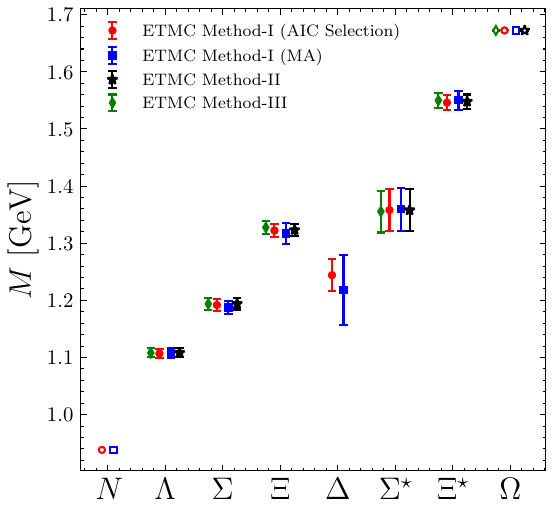}
    \caption{We show the masses of the $\Delta$ extracted with method-I and of the strange baryons calculated using all three methods. The red circles and blue squares denote results obtained with method-I i.e. from the fit having the largest probability and with averaging over all models, respectively. Black stars denote results obtained using method-II and green diamonds denote results using method-III. The empty symbols correspond to the baryons we used as input.}
    \label{fig:Strange_Methods}
\end{figure}

\begin{figure}[t!]
    \centering
    \includegraphics[width = 0.9\linewidth]{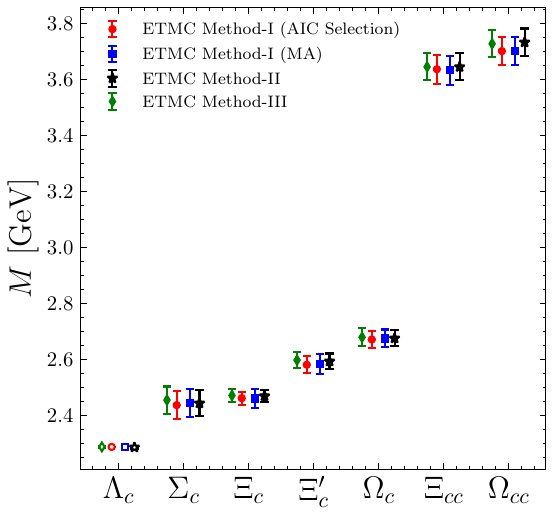}
    \caption{We show the masses for the spin-$1/2$  charmed baryons calculated using the three methods methods. The notation is the same as that of Fig.~\ref{fig:Strange_Methods}.}
    \label{fig:Charm1_Methods}
\end{figure}

\begin{figure}[t!]
    \centering
    \includegraphics[width = 0.9\linewidth]{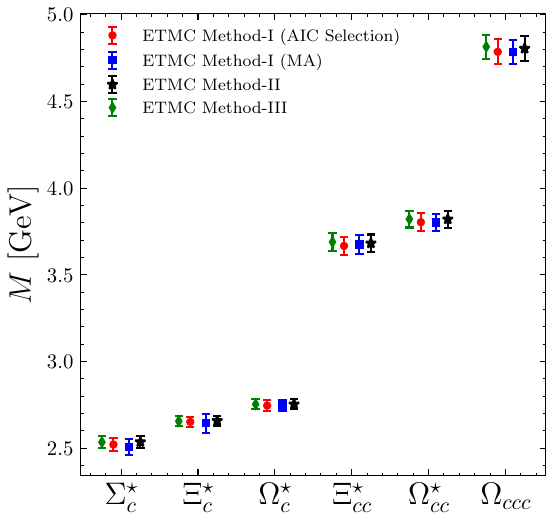}
    \caption{We show the masses for the spin-$3/2$ charmed baryons calculated using the three methods methods. The notation is the same as that of Fig.~\ref{fig:Strange_Methods}.}
    \label{fig:Charm2_Methods}
\end{figure}

A comparison among the continuum extracted results using the three methods is shown in Fig. \ref{fig:Strange_Methods} for the non-charmed baryons,  in Fig.~\ref{fig:Charm1_Methods} for the spin-1/2 charmed baryons, and in Fig.~\ref{fig:Charm2_Methods} for the spin-3/2 charmed baryons.  We observe very good agreement among values for all three methods. We thus opt to use as our final values the results obtained from method I with the model averaging, since within this method, one takes into account all ranges for the fits (models) eliminating a source of bias that may arise in choosing one particular fit. 

\begin{figure}[t!]
        \includegraphics[width = 0.9\linewidth]{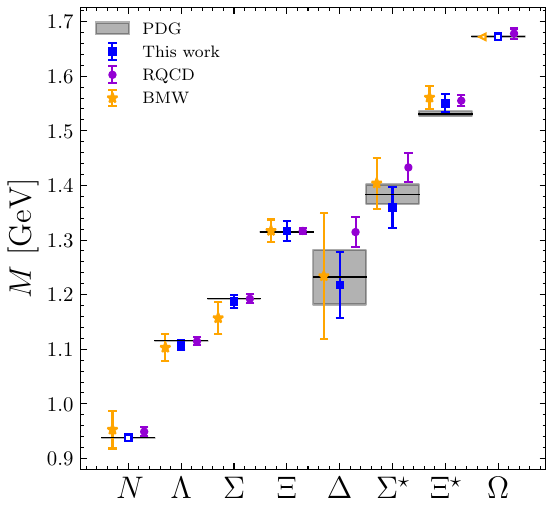}
        \caption{
        We compare our results for the masses of non-charmed baryons (top panel) (blue squares) with the results from RQCD ~\cite{RQCD:2022xux} (purple circles), and from BMW ~\cite{Durr:2008zz}(yellow stars). The horizontal bands denote experimental results from PDG~\cite{ParticleDataGroup:2022pth}. For the resonances $\Delta$, $\Sigma^\star$ and $\Xi^\star$ the width of the band is the resonance width.
        \label{fig:baryon_spectrum_strange}}
\end{figure}

\begin{figure}[t!]
    \centering
    \includegraphics[width = 0.9\linewidth]{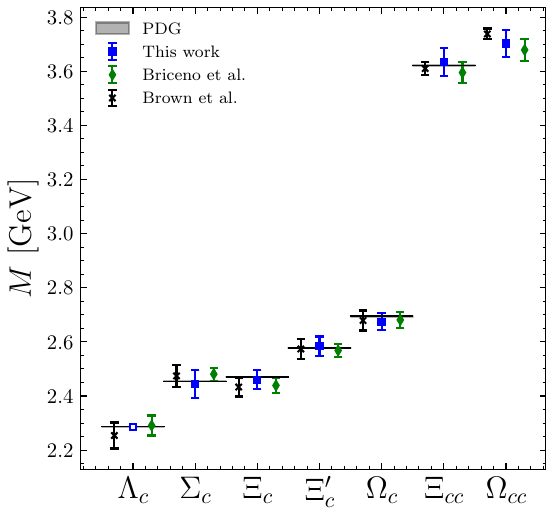}
    \includegraphics[width = 0.9\linewidth]{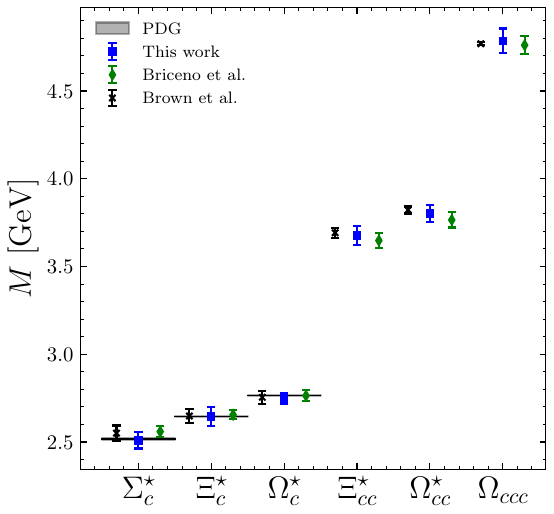}
    \caption{We compare our results for the spin-1/2 (top panel) and spin-3/2 (bottom panel) charmed baryons (blue squares) with the results from Ref.~\cite{Briceno:2012wt} (green diamonds) and Ref.~\cite{Durr:2008zz} (black crosses). We also compare with the same experimental results as in Fig.~\ref{fig:baryon_spectrum_strange}. For baryons that are resonances ($\Sigma_c^\star$, $\Xi_c^\star$, and $\Omega_c^\star$) we use the resonance width as the error.}
    \label{fig:baryon_spectrum_charm}
\end{figure}

The values for the baryons masses determined using method I and the model averaging are given in Table~\ref{tab:baryon_spectrum}.

\section{Comparison with other lattice QCD results}

Several collaborations have computed baryon masses, using ensembles that include larger than physical pion masses. This work is the first one to evaluate the low-lying baryon mass spectrum at the continuum limit using {\it only} physical point ensembles. We compare our results to other recent lattice QCD results. The BMW collaboration~\cite{Durr:2008zz} performed the first groundbreaking computation of baryon masses using unquenched gauge ensembles. They used several $N_f=2+1$ ensembles of clover-improved Wilson fermions simulated with pion masses down to $190\,\si{MeV}$ and with lattice spacings ranging from $0.125\,\si{fm}$ down to $0.065\,\si{fm}$. They performed simultaneous extrapolations to the physical pion mass and to the continuum limit and since they had several volumes they included finite-volume corrections in their analysis. Another collaboration~\cite{Briceno:2012wt} used a relativistic heavy-quark action for charm quarks and the clover-improved Wilson fermion action for the light and strange quarks in order to calculate the charmed baryon spectrum.
They used five ensembles with pion masses down to $220\,\si{MeV}$ with lattice spacings in the range $a \approx 0.12 - 0.06\,\si{fm}$ to extrapolate to the continuum limit and they performed simultaneous extrapolations to the physical light and strange quark masses.
In Ref.~\cite{Brown:2014ena}, a relativistic heavy-quark action is employed for the charm sector, while for the light and strange quarks, they used domain-wall fermions.
They used eight ensembles having two different lattice spacings of $(a \approx 0.11\text{ and } 0.085\,\si{fm})$ and of pion masses down to $227\,\si{MeV}$ to extrapolate to the physical point and to the continuum limit.

\begin{table*}[t!]
\centering
                \begin{tabular}{c|c|c|c|c|c}
                        \multicolumn{6}{c}{Strange baryons}\\
                        \hline
                        \multirow{2}{*}{$\Lambda(1.116)$} & \multirow{2}{*}{$\Sigma(1.193)$} & \multirow{2}{*}{$\Xi(1.314)$} & \multirow{2}{*}{$\Delta(1.232)$} & \multirow{2}{*}{$\Sigma^\star(1.385)$} & \multirow{2}{*}{$\Xi^\star(1.530)$}\\
                         & & & & & \\
                        1.1079(84) & 1.187(12) & 1.317(18) & 1.218(61) & 1.360(38) & 1.550(16)\\
                        \hline\hline
                        \multicolumn{6}{c}{}\\
                        \multicolumn{6}{c}{Spin-$1/2$ Charmed baryons}\\
                        \hline
                        \multirow{2}{*}{$\Sigma_c(2.455)$} & \multirow{2}{*}{$\Xi_c(2.470)$} & \multirow{2}{*}{$\Xi_c^\prime(2.578)$} & \multirow{2}{*}{$\Omega_c^0(2.695)$} & \multirow{2}{*}{$\Xi_{cc}(3.622)$} & \multirow{2}{*}{$\Omega_{cc}$}\\
                         & & & & & \\
                        2.443(51) & 2.460(35) & 2.584(36) & 2.675(32) & 3.634(51) & 3.703(51)\\
                        \hline\hline
                        \multicolumn{6}{c}{}\\
                        \multicolumn{6}{c}{Spin-$3/2$ Charmed baryons}\\
                        \hline
                        \multirow{2}{*}{$\Sigma_c^\star(2.520)$} & \multirow{2}{*}{$\Xi_c^\star(2.645)$} & \multirow{2}{*}{$\Omega_c^\star(2.770)$} & \multirow{2}{*}{$\Xi_{cc}^\star$} & \multirow{2}{*}{$\Omega_{cc}^\star$} & \multirow{2}{*}{$\Omega_{ccc}$}\\
                         & & & & & \\
                        2.509(46) & 2.645(55) & 2.747(31) & 3.676(55) & 3.803(50) & 4.785(71)\\
                        \hline\hline
                \end{tabular}
        \caption{Our final values for the low-lying baryon spectrum determined using method I with model averaging.
    The  experimental mass from the PDG~\cite{ParticleDataGroup:2022pth}, if known, is given in the parenthesis next to the baryon symbol in the first row.}
        \label{tab:baryon_spectrum}
\end{table*}

Very recently, the RQCD collaboration \cite{RQCD:2022xux} used 58 $N_f=2+1$ ensembles of clover-improved Wilson fermions with six different lattice spacings starting from $a = 0.10\,\si{fm}$ down to  $a \sim 0.04\,\si{fm}$.
These ensembles yielded pion masses ranging from $429\,\si{MeV}$ down to $127\,\si{MeV}$ and using them they performed a chiral extrapolation to the physical point, infinite volume, and continuum extrapolations to obtain the strange baryon spectrum.

In Figs.~\ref{fig:baryon_spectrum_strange} and \ref{fig:baryon_spectrum_charm}, we compare our results with those of the aforementioned collaborations as well as with experimental results \cite{ParticleDataGroup:2022pth}.

We observe excellent agreement between our results and previous lattice QCD results. They are also in agreement with the experimental values when available. We find that lattice QCD results on the mass of $\Xi^*$ favor a larger central value by about 20~MeV.

\section{Conclusions}

\begin{figure}[t!]
    \includegraphics[width = 0.9\linewidth]{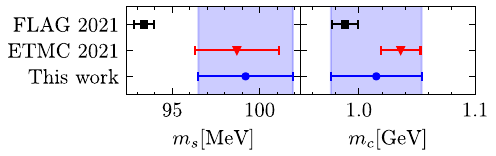}
    \caption{Results on the strange (left) and charm (right) quark masses (blue circles) compared to the FLAG21 average (black squares)~\cite{FlavourLatticeAveragingGroupFLAG:2021npn}  and our previous results (red triangles)~\cite{ExtendedTwistedMass:2021gbo}.\label{fig:quark}}
\end{figure}

Using three $N_f=2+1+1$ ensembles simulated with physical values of the quark masses, we perform for the first time directly at the physical pion mass the continuum limit for all low-lying baryon masses. The volume for these three ensembles is approximately the same with a spatial extent of 5.1 fm for the cB211.072.64 ensemble and  5.5 fm for the other two ensembles. The lattice spacings are determined using the nucleon mass after applying a correction for the small miss-tuning in the pion mass. 
We use the twisted mass fermion formulation with clover improvement and  Osterwalder-Seiler valence strange and charm quarks. We tune the valence quark masses using the physical masses of the $\Omega^-$ and the $\Lambda_c^+$   baryons, respectively.
The renormalized strange and charm quark masses that we find are 
$m_s(2~\text{GeV})=99.2(2.7)\,\si{MeV}$ and $m_c(3~\text{GeV})=1.015(39)\,\si{GeV}$, respectively in the $\overline{\text{MS}}$ scheme at the continuum limit. We compare these values in Fig.~\ref{fig:quark}. There is agreement with the values of our previous analysis~\cite{ExtendedTwistedMass:2021gbo} that did not include the  cD211.054.96 ensemble but included ensembles simulated at larger than physical pion mass that enabled us to include ensembles with $a\sim 0.093$~fm.   While there is agreement with the FLAG average~\cite{FlavourLatticeAveragingGroupFLAG:2021npn} for the charm quark mass, the one standard deviation tension for the strange quark mass persists.

Our values for the baryon masses are in agreement with experimental results. For the $\Xi^\star$ baryon, we find $M_{\Xi^\star} = 1.550(16)\,\si{GeV}$,  as compared to the experimental value of 1.530~GeV and consistent with other two lattice QCD determinations\cite{Durr:2008zz, RQCD:2022xux}. Although within about a standard deviation the lattice QCD results agree with the experimental value, all three results yield a larger central value as compared to what  PDG quotes.
 This overall agreement with the experimental results allows us to predict the unmeasured baryon masses of doubly and triply charmed baryons.
The values predicted for the masses of the doubly charmed $\Xi_{cc}^\star$, $\Omega_{cc}$ and $\Omega_{cc}^\star$  baryons are 3.676(55)GeV, 3.703(51)
GeV and 3.803(50)GeV, respectively, and for the triply charmed $\Omega_{ccc}$ baryon 4.785(71)GeV.

\section*{Acknowledgments}
C.A. acknowledges support by the project 3D-nucleon, id number EXCELLENCE/0421/0043, co-financed by the European Regional Development Fund and the Republic of Cyprus through the Research and Innovation Foundation. 
G.C. was funded by the projects  NextQCD, id EXCELLENCE/0918/0129, and NiceQuarks id EXCELLENCE/0421/0195, co-financed by the European Regional Development Fund and the Republic of Cyprus through the Research and Innovation Foundation. 
 S.B. is funded by the project QC4LGT, id number EXCELLENCE/0421/0019, co-financed by the European Regional Development Fund and the Republic of Cyprus through the Research and Innovation Foundation. S.B. also acknowledges funding by the EuroCC project (grant agreement No. 951740).
This project used computer time under the project \texttt{pr74yo} on the JUWELS Booster system at the J\"{u}lich Supercomputing Centre and on the Cyclone supercomputer at The Cyprus Institute.
The authors gratefully
acknowledge the Gauss Centre for Supercomputing e.V.
(www.gauss-centre.eu) for funding this project by providing computing time through the John von Neumann
Institute for Computing (NIC) on the GCS Supercomputer JUWELS-Booster at J\"{u}lich Supercomputing
Centre (JSC). The authors also acknowledge the Texas Advanced Computing Center (TACC) at The University of Texas at Austin for providing HPC resources that have contributed to the research results.

\bibliography{references}

\end{document}